%% file: deltashell.tex
\documentclass[aps,twocolumn,showpacs,preprintnumbers,amsmath,amssymb,nofootinbib,superscriptaddress,showkeys]{revtex4-1}

\usepackage{hyperref}
\usepackage{float}
\usepackage{epsfig}
\usepackage{graphicx}
\usepackage{mathptmx}      % use Times fonts if available on your TeX system
\usepackage{latexsym}
\usepackage{natbib}
\usepackage{longtable}
\usepackage{dcolumn}

\DeclareMathVersion{nxbold}
\SetSymbolFont{operators}{nxbold}{OT1}{cmr} {b}{n}
\SetSymbolFont{letters}  {nxbold}{OML}{cmm} {b}{it}
\SetSymbolFont{symbols}  {nxbold}{OMS}{cmsy}{b}{n}

\DeclareMathAlphabet{\mathcal}{OMS}{cmsy}{m}{n}
\begin{document}

\title{Coarse grained Potential analysis of neutron-proton and
  proton-proton scattering below pion production
  threshold}~\thanks{Supported by Spanish DGI (grant FIS2011-24149)
  and Junta de Andaluc{\'{\i}a} (grant FQM225).  R.N.P. is supported
  by a Mexican CONACYT grant.}

\author{R. Navarro P\'erez}
 \email{rnavarrop@ugr.es}
 \affiliation{Departamento de F\'{\i}sica At\'omica, Molecular y Nuclear \\
 and Instituto Carlos I de F{\'\i}sica Te\'orica y Computacional \\ 
 Universidad de Granada, E-18071 Granada, Spain.}
\author{J.E. Amaro}
 \email{amaro@ugr.es}
 \affiliation{Departamento de F\'{\i}sica At\'omica, Molecular y Nuclear \\
 and Instituto Carlos I de F{\'\i}sica Te\'orica y Computacional \\ 
 Universidad de Granada, E-18071 Granada, Spain.}
\author{E. Ruiz Arriola}
 \email{earriola@ugr.es} 
 \affiliation{Departamento de F\'{\i}sica At\'omica, Molecular y Nuclear \\
 and Instituto Carlos I de F{\'\i}sica Te\'orica y Computacional \\ 
 Universidad de Granada, E-18071 Granada, Spain.}

\date{\today}

\begin{abstract} 
 Using the $\delta$-shell representation we present a successful fit to
 neutron-proton and proton-proton scattering data below pion
 production threshold.  A detailed overview of the theory necessary to
 calculate observables with this potential is presented. A new data
 selection process is used to obtain the largest mutually consistent
 data base. The analysis includes data within the years 1950 to 2013. Using 46
 parameters we obtain $\chi^2/N_{\rm data} = 1.04$ with $N_{\rm
   data}=6713$ including normalization data.  Phase shifts with error
 bars are provided.
\end{abstract}


\maketitle

\section{Introduction}

The determination of the nucleon-nucleon (NN) interaction has played a
central role in nuclear physics~\cite{Machleidt:1989tm}. So far, the only direct way to
determine the interaction from first principles and in terms of the
underlying quark and gluon degrees of freedom is by means of lattice
QCD calculations which will eventually come to realistic scenarios
(see e.g.~\cite{Beane:2010em,Aoki:2011ep} and references therein). The
traditional alternative to those incipient calculations is to
determine a phenomenological interaction from a partial wave analysis
(PWA) of the highly abundant (about 8000) scattering data. Equally
important and helpful should be a credible determination of theoretical
uncertainties in the interaction and its propagation to nuclear
structure calculations, an aspect that applies both to lattice QCD and
PWA. The necessary condition to carry out such a program is to achieve
in any case a chi square per degree of freedom $\chi^2/{\rm d.o.f}
\lesssim 1$ description of all available NN data when confronted
either with the predictions of the lattice QCD or the fitted
phenomenological interaction. Before 1990 all fits determining
phenomenological potentials which were routinely used in nuclear
structure calculations did not lower the value $\chi^2 / {\rm d.o.f}
\sim 2$ (for a historic account see
e.g. \cite{Machleidt:1989tm}). According to well known statistical
principles this prevents to estimate the errors due to statistical
fluctuations of the data. Only in the mid 90's was it possible to
provide high quality fits to np and pp scattering data with
$\chi^2/{\rm d.o.f.} \lesssim 1$ mainly due to i) the scrupulous
inclusion of charge-dependence (CD) including vacuum polarization,
relativistic corrections, magnetic moment interactions, among other
effects and ii) a sound rejection criterium of $3\sigma$-inconsistent
data with the validating NN interaction. Along these lines several
parameterizations have been proposed to describe a continuously
increasing database of np and pp experimental scattering data bellow
pion production threshold
\cite{Stoks:1993tb,Stoks:1994wp,Wiringa:1994wb,Machleidt:2000ge,Gross:2008ps}
and even to energies as high as 3 GeV for pp and 1.3 GeV for np
\cite{Arndt:2007qn} (in this latter case all data are included in the
analysis and $\chi^2 /{\rm d.o.f.} \sim 1.6$).  In
reference~\cite{NavarroPerez2013138} we presented a $\delta$-shell
potential fitted to pseudo-data that consisted of the mean and
standard deviation of the np phasehifts given by the Nijmegen
PWA~\cite{Stoks:1993tb} and six other potentials with $\chi^2/{\rm
  d.o.f.} \lesssim 1$
\cite{Stoks:1994wp,Wiringa:1994wb,Machleidt:2000ge,Gross:2008ps} and
obtained an estimate of the systematic uncertainties of the NN
interaction. A new PWA to pp and np data including experiments till
2013 was presented in \cite{Perez:2013mwa} from which statistical
uncertainties were extracted.  Here we present the details of that
work, paying special atention to the fitting procedure and the data
selection process.

We note that with the total NN database, comprising about 8000
scattering data, none of the available post-1993 analyses yields an
acceptable fit, i.e.  $\chi^2/{\rm dof} \lesssim 1$, by itself to
their contemporary {\it complete} data base (for a discussion about
the pre-1993 situation see e.g.~\cite{Machleidt:1992uz}). A dedicated
look at the data base shows that there are experiments which measure
several observables in quite similar and/or overlapping kinematical
conditions. However, a closer inspection reveals that certain data are
mutually incompatible within statistical errors. Clearly, this implies
that at least one data set is incorrect. Of course, the possibility of
several data sets being incorrect should not be discarded {\it a
  priori}.  The key question is which data or data set should be kept
and which ones should be rejected. All analysis carried out so far
approach this issue from the point of view of the tension between the
data and the model used to analyze them. This obviously introduces a
biass in the choice of the database, which can be included as a source
of systematic errors.  The Nijmegen group fixed the data base from
that point of view in 1993~\cite{Stoks:1993tb} and the high quality
phenomenological NijmI, NijmII, Reid93 and AV18 potentials developped
thereafter~\cite{Stoks:1994wp,Wiringa:1994wb} use the same selected
data to perform the analysis. The CD-Bonn potential
analysis~\cite{Machleidt:2000ge} kept the same accepted data base and
applied the $3\sigma$ criterion for the new pp and np data published
between 1993 and 1999 but the possible tension between pre- and
post-1993 data was not considered.  The covariant spectator model was
used to analyze np scattering data~\cite{Gross:2008ps} resulting in a
new selection of compatible data.

Following an interesting suggestion made by Gross and
Stadler~\cite{Gross:2008ps} proposing a refined $3\sigma$-criterion,
in this work we analyze the tension among each pair of data sets of
experiments performed and published from 1950 till 2013. As a
consequence, a large fraction of otherwise rejected data is rescued
with a statistical significance and in a model independent way.
Although we use a specific representation of the unknown part of the
interaction, if an unbiassed analysis is carried out, all errors
should be of purely statistical origin and the particular
representation should not play any role. In this regard, our
motivation to upgrade the PWA in a statistical meaningfull fashion was
the realization~\cite{NavarroPerez:2012vr,Perez:2012kt} that
discrepancies among different high quality fits were larger than the
declared statistical uncertainties.

The paper is organized as follows, section \ref{sec:Potential} defines
the potential, as well as the parametrization of the short and
intermediate range parts by the $\delta$-shell representation and the
expressions that describe the long range part. In section
\ref{sec:Fitting} the fitting procedure is laid out.  Special
attention is given to review the on-shell scattering amplitude
computaion, especially the electro-magnetic part, which appears
scattered in several publications and we collect here for the benefit
of the unfamiliarized readers. Section \ref{sec:DataSelection} details
the improved selection data criterion to obtain the largest database
without incompatible data. Extensive tables of accepted and rejected
data are also given. Section \ref{sec:Results} presents the results
and includes a table of the fitted parameters in the operator
basis. After error propagation with the pertinent correlations as
encoded in the standard covariance matrix is made, we also provide the
low angular momentum partial waves phase shifts with statistical
errors. Conclusions and outlook for further implementations of the
$\delta$-shell potential are given in section
\ref{sec:Conclusions}. Finally the appendix includes expresions that
relate our fitting parameters with parameters in the AV14 operator
basis and details for integrating the Schro\"odinger equation with a
$\delta$-shell potential both with central and tensor terms.

\section{Description of the potential} 
\label{sec:Potential}

For our purposes the NN interaction can be decomposed into different
known pieces featuring understood physical effects and unknown
contributions which are constrained with the help of the currently
existing data. In this paper we only considered {\it published} np and
pp scattering data. To this end many possible functional forms have
been proposed. In our previous
works~\cite{NavarroPerez:2011fm,Perez:2012kt} we have motivated the
use of the $\delta$-shell representation, $V_{\rm DS}(r)$, which was
first introduced in the NN context by Aviles in
1973~\cite{Aviles:1973ee}.  It consists of a sum of $N$ Dirac delta
functions, each one centered around a concentration radius $r_i$ and
multiplied by a strength coefficient $V_i$
 \begin{equation}
  V_{\rm DS}(r) = \sum_{i=1}^{N} V_i \delta(r-r_i).
  \label{DeltaShellPotential}
 \end{equation}
The computational advantages of this representation in nuclear
structure calculations and uncertainty estimation have already been
stressed ~\cite{NavarroPerez:2011fm,Perez:2012kt,Perez:2013mwa}. Using
this representation it is posible to accurately describe the short and
intermediate range part of the nucleon-nucleon interaction by fixing
the distance between concentration radii at $\Delta r = r_{i+1} -
r_{i} = 0.6$ fm and determining the strength coefficients by a fit to
scattering data below pion production
threshold~\cite{Perez:2013mwa}. The long range part consists of the
well known charge dependent one pion exchange (OPE) potential and
electromagnetic (EM) interactions. In its complete form the potential
reads
\begin{eqnarray}
   V(r) &=& \sum_{n=1}^{21} O_n \left[\sum_{i=1}^N V_{i,n} \Delta r_i \delta(r-r_i) \right] \nonumber \\ &+& \Big[ V_{\rm OPE}(r) + V_{\rm EM}(r) \Big] \theta(r-r_c),
\label{eq:potential}
\end{eqnarray}
where $O_n$ are a set of operators. The first eighteen operators
correspond to the ones used on the AV18
potential~\cite{Wiringa:1994wb}, the remaining three new operators
incorporate further charge dependence and are defined in
Appendix~\ref{sec:transform}~\footnote{We note a typo in our previous
  work~\cite{Perez:2013mwa} where there appears 18 instead of the
  21.}. The distinction between intermediate and long range is made
explicit by the cut-off radius $r_c$, which turns out to be $3.0$ fm
since the interaction above that distance is correctly described by
OPE and EM terms only with no finite size effects. Smaller values of
$r_c$ were also considered but did not come out as optimal.

Even though the strength coefficients $V_{i,n}$ can be fitted directly
to the data, the partial wave decomposition of the potential
\begin{eqnarray}
V^{JS}_{l,l'}(r) = \frac{1}{2\mu_{\alpha\beta}}\sum_{i=1}^N  (\lambda_i)^{JS}_{l,l'} \delta(r-r_i)  \qquad r \le r_c,
\label{eq:ds-pot}
\end{eqnarray}
where $\mu_{\alpha\beta}$ is the reduced mass with $\alpha,\beta=n,p$,
allows to directly incorporate charge dependence in the $^1S_0$
parameters. 
% only by using the strength coefficients $(\lambda_i)^{JS}_{l,l'}$ as
%fitting parameters.  To reconstruct the complete potential, the
strength coefficients in the operator basis can be written as a linear
combination of the $\lambda_i$ coefficients of low angular momentum
partial waves as is shown in the appendix~\ref{sec:transform}. In
practice the potential can be parameterized by using only fifteen
independent partial waves, therefore only fifteen operators will have
independent strength coefficientes, the rest will be either fixed to
zero or will be linearly dependent on other operators coefficients.  A
good reason to use the lowest partial wave coefficients as primary
fitting parameters is that correlations among different partial waves
turn out to be much smaller than the correlations between the operator
coefficients. Note that we are just making a change of basis, but the
coefficients of the higher partial waves are calculated by
constructing the complete potential and decomposing it into the
corresponding partial waves (see appendix~\ref{sec:transform}). The
resulting partial wave coeeficients were displayed in our previous
work~\cite{Perez:2013mwa}. Here we will show the equivalent results
for the operator coefficients (see Table~\ref{tab:Fits} below).

The charge dependent OPE potential in the long range part of the
interaction is the same as the one used by the Nijmegen group on their
1993 partial wave analisys~\cite{Stoks:1993tb} and reads
\begin{equation}
 V_{m, \rm OPE}(r) = f^2\left(\frac{m}{m_{\pi^\pm}}\right)^2\frac{1}{3}m\left[Y_m(r){\mathbf \sigma}_1\cdot\mathbf{\sigma}_2 + T_m(r)S_{1,2} \right]
\end{equation}
being $f$ the pion coupling constant, ${\mathbf \sigma}_1$ and ${\mathbf
\sigma}_2$ the single nucelon Pauli matrices, $S_{1,2}$ the tensor
operator, $Y_m(r)$ and $T_m(r)$ the usual Yukawa and tensor functions,
\begin{eqnarray}
 Y_m(r) &=& \frac{e^{-m r}}{m r}, \nonumber \\
 T_m(r) &=& \left(1+ \frac{3}{mr} + \frac{3}{(mr)^2} \right)\frac{e^{-m r}}{m r}. 
 \label{eq:Yukawa}
\end{eqnarray}
Charge dependence is introduced by the difference between
the charged $m_{\pi^\pm}$ and neutral $m_{\pi^0}$ pion mass by setting
\begin{eqnarray}
 V_{{\rm OPE},pp}(r) &=& V_{m_{\pi^0},\rm OPE}(r), \nonumber \\
 V_{{\rm OPE},np}(r) &=& -V_{m_{\pi^0},\rm OPE}(r)+ (-)^{(T+1)}2V_{m_{\pi^\pm},\rm OPE}(r).
 \label{eq:BreakIsospinOPE}
\end{eqnarray}

The neutron-proton electromagnetic potential includes only a magnetic
moment interaction
 \begin{equation}
V_{\rm EM, np}(r) = V_{\rm MM, np}(r) = -\frac{\alpha \mu_n}{2M_{n} r^3}  \left( \frac{\mu_{p}S_{1,2}}{2 M_p}  + \frac{{\bf L}\!\cdot\!{\bf S}}{\mu_{np}}  \right),
 \end{equation}
 where $\mu_n$ and $\mu_p$ are the nuetron and proton magnetic moments,
 $M_n$ the neutron mass, $M_p$ the proton one and ${\bf L}\!\cdot\!{\bf
 S}$ is the spin orbit operator. The EM terms in the proton-proton
 channel include one and two photon exchange, vacuum polarization and
 magnetic moment,
\begin{equation}
 V_{\rm EM, pp}(r) = V_{\rm C1}(r) + V_{\rm C2}(r) + V_{\rm VP}(r) + V_{\rm MM, pp}(r)
\end{equation}
where
 \begin{eqnarray}
  &V&_{\rm C1}(r) = \frac{\alpha'}{r} \ ,     \\
  &V&_{\rm C2}(r) = -\frac{\alpha\alpha'}{M_{p} r^2} \ ,  \\
  &V&_{\rm VP}(r) = \frac{2\alpha\alpha'}{3\pi r} \int^{\infty}_{1} e^{-2m_{e}rx}\left(1+\frac{1}{2x^{2}}\right)
            \frac{\sqrt{x^{2}-1}}{x^{2}}dx  \ ,                 \\
%            \frac{(x^{2}-1)^{1/2}}{x^{2}} \ ,                 \\
&V&_{\rm MM, pp}(r) = -\frac{\alpha}{4M^{2}_{p} r^3} \left[
                 \mu^{2}_{p}S_{1,2}  + 2(4\mu_{p}-1){\bf L}\!\cdot\!{\bf S}  \right].
\end{eqnarray}  
Note that these potentials are {\it only} used above $r_c = 3 {\rm
  fm}$ and thus form factors accounting for the finite size of the
nucleon can be set to one.  Energy dependence is present through the
parameter
 \begin{equation}
  \alpha'= \alpha\frac{1+2k^2/M_p^2}{\sqrt{1+k^2/M_p^2}},
 \end{equation}
 where $k$ is the center of mass momentum and $\alpha$ the fine
 structure constant. Table \ref{tab:Constants} lists the values used for
 the fundamental constants in this work's calculations.

\begin{table}
 \caption{\label{tab:Constants} Values of fundamental constants used.}
 \begin{ruledtabular}
 \begin{tabular*}{\columnwidth}{@{\extracolsep{\fill}} c D{.}{.}{3.7} l}
   Constant     & \multicolumn{1}{c}{Value}      & Units      \\
  \hline  
   $\hbar c$     & 197.327053 & MeV fm    \\
   $m_{\pi^0}$   & 134.9739   & Mev$/c^2$   \\
   $m_{\pi^\pm}$ & 139.5675   & Mev$/c^2$   \\
   $M_{p}$       & 938.27231  & Mev$/c^2$   \\
   $M_{n}$       & 939.56563  & Mev$/c^2$   \\
   $m_e  $       & 0.510999   & Mev$/c^2$   \\
   $\alpha^{-1}$ & 137.035989 &           \\
   $f^2$         & 0.075      &           \\
   $\mu_p$       & 2.7928474  & $\mu_0$     \\
   $\mu_n$       &-1.9130427  & $\mu_0$     \\
 \end{tabular*}
 \end{ruledtabular}
\end{table}

Even though the contribution of all non Coulomb electromagnetic terms
to the non central partial wave phaseshifts is rather small when
compared to the $V_{\rm C1}$ and $V_{\rm OPE}$ ones, their inclusion
is crucial to accurately describe the scattering amplitude. Also the
vacuum polarization contribution is needed for the proper calculation
of low energy observables. For these reasons, in this work the
potential in the $^1S_0$ partial wave includes all EM terms listed
previously, while the rest has only the $V_{\rm C1}$ one. Still, the
electromagnetic scattering amplitude is constructed with all terms
explicitly, as is shown below.

The Coulomb effects in the short range part of the interaction are
included by a coarse grained representation, instead of simply extending
$V_{\rm C1}$ bellow $r_c$, in order to keep the advantage of having only
a few interaction radii $r_i$ in that region. This coarse graining is
obtained by looking for a $\delta$-shell represention of the
interaction, i.e. $\bar V_{\rm C1} (r) = \sum_n V_i^C \Delta r_i \delta
(r-r_i) + \theta (r-r_c) V_{\rm C1} (r) $, where the $V_i^C$ are
determined by reproducing the Coulomb scattering amplitude to
high-precision and are not changed in the fitting process. The first
line of table~\ref{tab:Fits} shows the corresponding $\delta$-shell
parameters $V_i^C$.

\section{Fitting procedure}
\label{sec:Fitting}

The determination of the $V_{i,n}$ coefficients in
Eq.~(\ref{eq:potential}) is made through a partial wave decomposition
of the potential to calculate observables and reproduce experimental
data. Our fitting procedure consists of using the
$(\lambda_i)^{JS}_{l,l'}$ parameters to calculate partial wave
phaseshifts, summing those phaseshifts to obtain the scattering
amplitude $M$, extracting observables from $M$, comparing observables
with experimental data by a merit function and minimizing such
function with respect of the fitting parameters. In theory such a
procedure requires a sum of an infinite number of partial waves in the
complete scattering amplitude, but all high angular momentum partial
waves in the potential can be written as linear combinations of the
low angular momentum ones by means of the relation between the later
and the operator basis (see App.~\ref{sec:transform}). This allows to
use only a few $(\lambda_i)^{JS}_{l,l'}$ coefficients as independent
fitting parameters. Also, the phase shifts of very high angular
momentum partial waves are mostly determined by the long range part of
the interaction and their contribution to the scattering amplitude is
limited by the centrifugal barrier, therefore in practice a limited
number of partial waves is needed and summing up to $J_{\rm max} =20\,
$ proofs sufficiently accurate to compute the strong scattering
amplitude below pion production threshold.

Phase shifts are calculated by integrating Schr\"odinger's equation,
the details of such calculation with the $\delta$-shell potential are
given in App.~\ref{sec:sch}. The next subsection
reproduces and outlines the expresions necessary to calculate the
nuclear and electro-magnetic scattering amplitudes.
Reference~\cite{bystricky1978formalism} has an exhaustive list of
observables that can be extracted from different parametrizations of
$M$. The calculation of the merit function $\chi^2$ is explained in
subsection~\ref{subsec:chisquare}, special attention is given to the
treatment of systematic uncertainties from the experimental data.

In any fit we have {\it always} constrained the potential to reproduce
the deuteron binding energy to its experimental value $B=2.224575(9)$
MeV as well as the np $^1S_0$ scattering length to $\alpha_{^1S_0}=-23.74(2)$ fm.

\subsection{Scattering Amplitude}

The on-shell scattering amplitude $M$ can be expressed in terms of five
complex quantities, several parametrizations exist for this porpuse and
for definiteness we choose the wolfenstein parameters where
\begin{eqnarray}
   M(\mathbf{k}_f,\mathbf{k}_i) &=& a + m (\mathbf{\sigma}_1,\mathbf{n})(\mathbf{\sigma}_2,\mathbf{n}) 
                  + (g-h)(\mathbf{\sigma}_1,\mathbf{m})(\mathbf{\sigma}_2,\mathbf{m}) \nonumber \\
                  & &+ (g+h)(\mathbf{\sigma}_1,\mathbf{l})(\mathbf{\sigma}_2,\mathbf{l})  + c(\mathbf{\sigma}_1+\mathbf{\sigma}_2,\mathbf{n}) \, ,  
\end{eqnarray}
where  $\mathbf{l}$, $\mathbf{m}$, $\mathbf{n}$ are three
unitary orthogonal vectors along the directions of
$\mathbf{k}_f+\mathbf{k}_i$, $\mathbf{k}_f-\mathbf{k}_i$ and
$\mathbf{k}_i \wedge \mathbf{k}_f$ and $\mathbf{k}_f$, $\mathbf{k}_i$
are the final and initial relative nucleon momenta respectively. The
parameters $a,m,g,h,c$ depend on the scattering angle $\theta$ and $k$,
also any scattering observable in our database can be written in terms
of them
\cite{bystricky1978formalism, bystricky1987nucleon}.

The partial wave decomposition of the $M^s_{m_s',m_s}$ matrix elements
due to a certain interaction is
\begin{eqnarray}
 M^s_{m_s',m_s}(\theta) &=&  \frac{1}{2ik} \sum_{J,l',l}\sqrt{4\pi(2l+1)}Y^{l'}_{m_s'-m_s}(\theta,0) \nonumber \\
      &\times& C^{l',s,J}_{m_s-m_s',m_s',m_s}i^{l-l'}(S^{J,s}_{l,l'}-\delta_{l',l}) C^{l,s,J}_{0,m_s,m_s},
 \label{eq:MmatrixPartialWaves}
\end{eqnarray}
where $C^{l,s,J}_{m_l,m_s,m_J}$ is a Clebsch-Gordan coefficient,
$Y^{l}_{m}(\theta,\phi)$ the spherical harmonic, $\delta_{l,l'}$ a
Kronecker delta and $S^{J,s}_{l,l'}$ are the $S$ matrix elements with
the corresponding phaseshifts of such interaction. Denoting the phase
shifts as $\delta^{J,s}_{l,l'}$, for the singlet ($s=0$, $l = l'= J$)
and triplet uncoupled ($s=1$, $l=l'=J$) channels the $S$ matrix is
simply $e^{2i\delta^{J,s}_{l,l}}$, in the triplet coupled channel
($s=1$, $l=J\pm1$, $l'=J\pm1$) it reads
\begin{equation}
S^J = \left( 
\begin{array}{c c}
 e^{2i\delta^{J,1}_{J-1}} \cos{2 \epsilon_J} & ie^{i(\delta^{J,1}_{J-1}+\delta^{J,1}_{J+1})} \sin{2 \epsilon_J} \\
 ie^{i(\delta^{J,1}_{J-1}+\delta^{J,1}_{J+1})} \sin{2 \epsilon_J} & e^{2i\delta^{J,1}_{J+1}} \cos{2 \epsilon_J}
\end{array}
\right),
\end{equation}
with $\epsilon_J$ the mixing angle. The scattering amplitude has a
contribution for every term considered in the potential, this allows to
separate $M$ in a part due to the nuclear interaction and another coming
from the EM terms,
\begin{equation}
 M = M_{\rm EM} + M_{\rm N}.
\end{equation}
The pp and np electro magnetic amplitudes read
\begin{eqnarray}
M_{\rm EM, pp} &=& M_{\rm C1} + M_{\rm C2} + M_{\rm VP} + M_{\rm MM, pp}, \\
M_{\rm EM, np} &=& M_{\rm MM, np}.
\end{eqnarray}

Given the finite range nature of the nuclear interaction, $M_{\rm N}$
has a fast convergence when summing over partial waves and allows a
rapid calculation every time the fitting parameters are varied during
the fitting procedure. Meanwhile, the $M_{\rm EM}$ part of a pp
scattering has a slow convergence due to the interplay among different
long range contributions. Actually, $M_{\rm C2}$ and $M_{\rm MM,pp}$
require sumations up to $l=1000$. Fortunately, since $M_{\rm EM}$ does
not depend on the fitting parameters it only has to be calculated
once and stored. 

The expresions to calculate every part of the pp electromagnetic
scattering amplitude are well
known~\cite{Stoks:1993tb,Durand:1957zz,Stoks:1990us} and we reproduce
them here for completeness. The Coulomb scattering amplitude is given by
\begin{eqnarray}
 f_{{\rm C1 },k}(\theta) &=& \frac{1}{2ik}\sum_l{(2l+1)\left[e^{2i(\sigma_l-\sigma_0)}-1 \right]P_l(\theta)} \nonumber \\
  & = & - \frac{\eta}{k}\frac{e^{-i\eta\ln\frac{1}{2}(1-\cos\theta)}}{1-\cos{\theta}},
\end{eqnarray}
where $P_l(\theta)$ are the Legendre polynomials, $ \eta = \alpha' M_p
/(2k)$, and the Coulomb phaseshifts are calculated with
$\sigma_l=\arg\Gamma(l+1+i\eta)$. 

Since the two photon exchange potential $V_{\rm C2}$ has a $1/r^2$
dependence it can be absorved into the centrifugal barrier of the radial
Shcr\"onger's equation and the later is solved analytically using Coulomb
wave functions of noninteger $l$. This procedure leads to the $V_{\rm
C2}$ phasesifts
\begin{equation}
 \rho_l = \sigma_\lambda - \sigma_l + \frac{(l-\lambda)\pi}{2},
\end{equation}
where $\lambda$ is obtained by solving $\lambda(\lambda-1) = l(l+1) -
\alpha \alpha'$. Now the amplitude can be calculated with
\begin{equation}
f_{{\rm C2 },k} (\theta) = \frac{1}{2ik} \sum_l{(2l+1)e^{2i(\sigma_l-\sigma0)}\left[e^{2i\rho_l}-1\right]P_l(\theta)}.
\label{eq:C2ScatteringAmp}
\end{equation}

A similar expresion as the one in equation (\ref{eq:C2ScatteringAmp})
describes the vacumm polarization scatering amplitude replacing the
$\rho_l$ phashifts for the VP ones, which are usually denoted by
$\tau_l$. Since the values for $\tau_l$ are rather small, even for large
values of $k$ and $l$, the approximation
\begin{equation}
f_{{\rm VP },k} (\theta) = \frac{1}{k} \sum_l{(2l+1)e^{2i(\sigma_l-\sigma0)} \tau_l P_l(\theta)}
\label{eq:VPScatteringAmp}
\end{equation}
is a good starting point to calculate $f_{{\rm VP },k}$ as it can be
expanded by a series in powers of $\eta$ where $f_{\rm VP}=f^{(0)}_{\rm
VP }+f^{(1)}_{\rm VP }+f^{(2)}_{\rm VP }+\dots$ The leading order is
obtained using the first Born aproximation and expressed as
\begin{equation}
 f_{{\rm VP},k}^{(0)}(\theta) = - \frac{\alpha}{3 k \pi} \eta \frac{F(k,\theta)}{1- \cos{\theta}},
\end{equation}
where
\begin{eqnarray}
 F(k,\theta) &=& -\frac{5}{3} + X + \sqrt{1+X}\left(1-\frac{X}{2}\right) \nonumber \\
             &\times& \ln{\left[\frac{(1+X)^{1/2}+1}{(1+X)^{1/2}-1}\right]},
\end{eqnarray}
with $X = 2m_e^2/ [k^2(1-\cos\theta)]$. The real part of the subleading
order term can be computed with
\begin{eqnarray}
 {\rm Re}\left[ f_{{\rm VP},k}^{(1)}(\theta)\right] &=& \frac{4\alpha}{3 k \pi(1-\cos\theta)}\eta^2\left(\frac{1-\cos\theta}{1+\cos\theta}\right)^{1/2} \nonumber \\
 &\times& \left[\tan^{-1}\left(\frac{1+\cos\theta}{1-\cos\theta}\right)^{1/2} \right. \nonumber \\ 
 && \left. -\tan^{-1}\left(\frac{m_e^2}{k^2}\frac{1+\cos\theta}{1-\cos\theta}\right)^{1/2} \right]
\end{eqnarray}
and a good approximation for the corresponding imaginary part is
\begin{eqnarray}
 {\rm Im}\left[ f_{{\rm VP},k}^{(1)}(\theta)\right] &\approx& \frac{\alpha}{3 k \pi(1-\cos\theta)}\eta^2\left[\ln\left(\frac{1}{X}\right)\right] \nonumber \\
 &\times& \left[\ln\left(\frac{k}{m_e}\right) - \frac{3}{2} \ln\left(\frac{2}{1-\cos \theta}\right) \right].
\end{eqnarray}
An expansion up to this order has been found to be accurate enough to
describe $f_{{\rm VP},k}$ for the energy range discussed in this work.

The treatment of identical particles in a pp scattering reaction
requires the antisymmetrization of the $M_{m_s',m_s}^s$ matrix elements,
this is easily done by 
\begin{equation}
 M_{{\rm X}m_s',m_s}^{s} = [f_{{\rm X },k}(\theta) + (-)^s f_{{\rm X },k}(\pi - \theta)] \delta_{m_s',m_s}
 \label{eq:AntisymmetricMX}
\end{equation}
where ${\rm X = C1, C2, VP}$.

For the magnetic moment pp amplitude it is necessary to calculate the
partial wave $K$ matrix which is defined by $ S - 1 = 2 i K(1-i
K)^{-1}$. Since $V_{\rm MM,pp}$ is proportional to the spin-orbit and
tensor operator there is no contribution to the spin singlet channel and
$(K_{\rm MM,pp})^{J,0}_{l,l} = 0$, the spin triplet channel elements are
given by
\begin{eqnarray}
 &(&K_{\rm MM,pp})^{J=l,1}_{l,l} = -M_p k(2f_T-f_{LS})I_{l,l},  \nonumber \\
 &(&K_{\rm MM,pp})^{J=l+1,1}_{l,l}   = -M_p k \left(-\frac{2l}{2l+3}f_T-lf_{LS} \right)I_{l,l}, \nonumber \\
 &(&K_{\rm MM,pp})^{J=l+1,1}_{l+2,l+2} = -M_p k \left(-\frac{2l+6}{2l+3}f_T-(l+3)f_{LS} \right)I_{l+2,l+2}, \nonumber \\
 &(&K_{\rm MM,pp})^{J=l+1,1}_{l,l+2} = -M_p k \left(6\frac{\sqrt{(l+1)(l+2)}}{2l+3}f_T \right)I_{l,l+2},
 \label{eq:KMMmatrix}
\end{eqnarray}
where $f_T$ and $f_{LS}$ are the coefficients of the tensor
and spin-orbit operators in the potential, i.e.
\begin{eqnarray}
 f_T &=& - \frac{\alpha \mu_p^2}{4 M_p^2}, \nonumber \\
 f_{LS} &=& -\frac{\alpha (4 \mu_p - 1)}{2 M_p^2},
\end{eqnarray}
the $I_{l,l'}$ terms are integrals of the $1/r^3$ dependence with
Coulomb wave functions and are given by
\begin{eqnarray}
 I_{l,l} &=& \frac{1}{2l(l+1)} \nonumber \\ 
  &+& \frac{1-\pi\eta+\pi\eta\coth(\pi\eta)-2\eta^2\sum_{n=0}^l(n^2+\eta^2)^{-1}}{2l(l+1)(2l+1)}, \nonumber \\
  I_{l,l+2} &=& \frac{1}{6} \left|l+1+i\eta \right|^{-1} \left|l+2+i\eta \right|^{-1}.
\end{eqnarray}
In principle the set of equations in (\ref{eq:KMMmatrix}) allows to
calculate $S_{\rm MM,pp}$ and use the later to obtain the $M_{\rm
MM,pp}$ matrix elements via the partial wave decomposition of eq.
(\ref{eq:MmatrixPartialWaves}), unfortunately the numerical effort to
reach convergence by summing over $J$, $l$ and $l'$ is too big to be
practical. Using the approximation $S_{\rm MM,pp} - 1 \approx 2 i K_{\rm
MM,pp}$ gives rise to a contribution
\begin{equation}
 Z_{LS} = - \frac{M_p}{\sqrt{2}}f_{LS}\sum_{{\rm odd} \ l} e^{2i(\sigma_l-\sigma_0)}\frac{2l+1}{l(l+1)}P_l(\theta)
 \label{eq:ZLSinMM}
\end{equation}
to the $M^1_{1,0}$ matrix element and the same with a minus sign to the
$M^1_{0,1}$ one. Fortunately, this series can be calculated analytically
with
\begin{eqnarray}
 Z_{LS} = - \frac{M_pf_{LS}}{\sin\theta\sqrt{2}}&&\left(e^{-i\eta\ln(1/2)(1-\cos\theta)} \right. \nonumber \\
                                                &&+\left.e^{-i\eta\ln(1/2)(1+\cos\theta)}-1 \right).
\end{eqnarray}
The use of this result significantly improves the convergence rate of
$M_{\rm MM,pp}$.

The neutron-proton EM amplitude can be expressed in terms of the
Wolfenstein-like parameters
$a,b,c,d,e$~\cite{Lechanoine:1979xj,LaFrance:1981ar,Stoks:1990us}
usually known as the Saclay parameters, and we reproduce this result for
completeness as well:
\begin{widetext}
\begin{align}
 a_{\rm EM,np}(s,t) = & \ \frac{\alpha}{t\sqrt{s}} \left\{\left(F_1^nF_1^p+tF_2^nF_2^p\right) \left[s- M_n^2 - M_p^2 + \frac{t^{\vphantom{2}}}{8sk^2} \left\{ \left[s- \left(M_n+M_p \right)^2 \right] \left[3s- \left(M_n-M_p \right)^2 \right] \right. \right. \right. \nonumber \\
  & \  + \left. \left. \left. 2\left[s- \left(M_n-M_p \right)^2 \right]\left(\sqrt{s}- M_n-M_p  \right)^2 \right\} + \frac{t^2}{16sk^4}\left[s- \left(M_n-M_p \right)^2 \right]\left(\sqrt{s}- M_n-M_p  \right)^2 \right] \right. \nonumber \\
  & \  + \left. \left(F_1^nF_2^p+F_2^nF_1^p\right)t \left[2 \sqrt{s} -M_n -M_p + \frac{t}{2k^2}\left(\sqrt{s}-M_n-M_p\right)  \right] \vphantom{\frac{t^2}{k^2}}  \right\}, \nonumber  \\
   b_{\rm EM,np}(s,t) = & \ \frac{\alpha}{t\sqrt{s}} \left[\left(F_1^nF_1^p-tF_2^nF_2^p\right) \left\{s- M_n^2 - M_p^2 + \frac{t}{8sk^2}  \left[s+ \left(M_n-M_p \right)^2 \right] \left[s- \left(M_n+M_p \right)^2 \right] \right\} \right. \nonumber \\
  & \ + \left. \left(F_1^nF_2^p-F_2^nF_1^p\right)t \left(M_m-M_p\right) \vphantom{\frac{t}{k^2}} \right] , \nonumber \\
  c_{\rm EM,np}(s,t) = & \  \frac{\alpha}{2\sqrt{s}}\left(F_1^n+2M_nF_2^n\right)\left(F_1^p+2M_pF_2^p\right), \nonumber \\
  d_{\rm EM,np}(s,t) = & \ - c(s,t), \nonumber \\
  e_{\rm EM,np}(s,t) = &  \ - i\frac{\alpha\sin\theta}{t\sqrt{s}} \left[\left(F_1^nF_1^p+tF_2^nF_2^p\right) \left\{s- M_n^2 - M_p^2 - \frac{M_n+M_p}{2\sqrt{s}}  \left[s+ \left(M_n-M_p \right)^2 \right] + \frac{\sqrt{s}-M_n-M_p}{\sqrt{s}+M_n+M_p} \frac{t}{2}  \right\} \right. \nonumber \\
  & \ + \left. \left(F_1^nF_2^p+F_2^nF_1^p\right)\left[2k^2\sqrt{s}+t\left(\sqrt{s}-M_m-M_p\right)\right]  \vphantom{\frac{M_n+M_p}{2\sqrt{s}}} \right],
  \end{align}
\end{widetext} 

where $s$ and $t$ are the Mandelstam invariants~\cite{Mandelstam:1958xc}
and can be calculated by $k^2 = [s-(M_n+M_p)^2][s-(M_n-M_p)^2]/4s$ and
$t = -2k^2(1-\cos\theta)$. $F_1$ and $F_2$ are the Dirac and Pauli form
factors, which on the point-particle approximation read
\begin{equation}
 F_1^p = 1, \ \ F_1^n = 0, \ \ F_2^p = \frac{\mu_p-1}{2M_p}, \ \ F_2^n = \frac{\mu_n}{2M_n}.
\end{equation}
The transformation between the Wolfensetein and Saclay parametrizations
can be found in \cite{bystricky1978formalism}.

One important remark has to be made about the $S$ matrix and the
phase-shifts that describe it. The nuclear phase-shifts presented in
this work are extracted with respect to the EM wave functions, as this
is also the case for the many other phase-shift analysis and
potentials in the literature
\cite{Perez:2013mwa,Wiringa:1994wb,Stoks:1993tb,Machleidt:2000ge,Gross:2008ps,Arndt:2007qn}.
For this reason, in the pp channel the $S_{\rm N}$ matrix in equation
(\ref{eq:MmatrixPartialWaves}) one has to make the replacement
\begin{equation}
 S_{\rm N} - 1 \rightarrow e^{i(\sigma_l+\rho_l+\tau_l)}(S_{\rm MM,pp})^{\frac{1}{2}}(S_{\rm N} - 1)(S_{\rm MM,pp})^{\frac{1}{2}}e^{i(\tau_l+\rho_l+\sigma_l)}.
\end{equation}
The np channel has a similar correction due to the magnetic moment
potential but its contribution is rather small and given the larger
uncertainties of the data the effect is not statistically significant,
therefore we do not include it in our calculations. Also, it should be
noted that the $\rho_l$ and $\tau_l$ phaseshifts, as well as the $K$
matrix elements of equation (\ref{eq:KMMmatrix}), are calculated with
respect to the Coulomb wave functions, which gives rise to the
$e^{2 i(\sigma_l-\sigma_0)}$ term in equations (\ref{eq:C2ScatteringAmp}),
(\ref{eq:VPScatteringAmp}) and (\ref{eq:ZLSinMM}).

\subsection{Calculation and minimization of the merit function $\chi^2$} 
\label{subsec:chisquare}

 With the scattering amplitude described by the Wolfenstein parameters
 it is possible to calculate observables for any scattering angle
 $\theta$ and center of mass momentum $k$, calculate a chi square
 merit function $\chi^2$ to asses the ability of the $\delta$-shell
 potential to reproduce the experimental data in our database, and
 adjust the $(\lambda)^{JS}_{l,l'}$ parameters by a least squares
 fitting to minimize $\chi^2$. The data are grouped by experiments and
 most of those measure an observable at a single laboratory frame
 energy $E_{\rm LAB}$ at different scattering angles, only a few of
 them are measured at a fixed angle for different values of $E_{\rm
   LAB}$ and the total cross section experiments include measurements
 also at different laboratory energies since there is no scattering
 angle involved.
 
 Every experimental data set can be subject to a known and common
 systematic uncertainty (normalized data), an arbitrarily large
 systematic uncertainty (floated data) or no systematic uncertainty at
 all (absolute data), this is recorded by the experimentalist everytime
 a set of measurements is made. In all three cases the merit function
 $\chi^2_t$ of a single data set is given by
 \begin{equation}
  \chi^2_t = \sum_{i=1}^n \frac{[o_i(k_i,\theta_i)/Z - t_i(k_i,\theta_i,(\lambda)^{JS}_{l,l'})]^2}{[\delta o_i(k_i,\theta_i)/Z]^2} + \frac{(1-1/Z)^2}{(\delta_{\rm sys}/Z)^2},
  \label{eq:ChiSquareSets}
 \end{equation}
where $o_i$ and $\delta o_i$ are the experimental value of and
observable and the corresponding statistical uncertainty at point $i$,
$t_i$ the theoretical value, $\delta_{\rm sys}$ the systematic
uncertainty of the experiment and $Z$ is a scaling factor. The last term
in equation (\ref{eq:ChiSquareSets}) is usually denoted as $\chi^2_{\rm
sys}$. Absolute data have $\delta_{\rm sys} = 0$ and are not scaled
($Z=1$). The correct value of $Z$ for normalized and floated data is
obtained by minimizing $\chi^2_t$ with respect to $Z$, this leads to 
\begin{equation}
 Z = \left(\sum_{i=1}^n \frac{o_i t_i}{\delta o_i^2} + \frac{1}{\delta_{\rm sys}^2} \right) \left/ \left(\sum_{i=1}^n \frac{t_i^2}{\delta o_i^2} + \frac{1}{\delta_{\rm sys}^2}\right) \right. ,
 \label{eq:ScalingZ}
\end{equation}
where the $k$, $\theta$ and $\lambda$ dependence has been
omitted. Since floated data have an arbitrarily large and common
systematic uncertainty, this type of data use equation
(\ref{eq:ScalingZ}) with $\delta_{\rm sys} = \infty$ so $\chi^2_{\rm
  sys} = 0 $. For normalized data the value of $\delta_{\rm sys}$ is
given by experimentalists, in most cases $Z\neq 1$, therefore
$\chi^2_{\rm sys} \neq 0 $ and the normalization is counted as an
extra data point. In some normalized data sets the systematic
uncertainty can give a rather large contribution to $\chi^2_t$,
probably due to a underestimation of $\delta_{\rm sys}$. To correct
for this understimation, if a data set has $\chi^2_{\rm sys} > 9$ we
float this data and no extra normalization data is counted, this is in
line with the $3 \sigma$ criterion which will be explained bellow
. Finally, the total $\chi^2$ is simply the sum of $\chi^2_t$ of every
np and pp data set.

To minimize the merit function $\chi^2$ with respect to the fitting
parameters $(\lambda)^{JS}_{l,l'}$ we use the Levenberg-Marquardt
method. This method requires the calculation of derivatives of every
calculated observable $t_i$ with respect to all the fitting parameters
to construct an approximation to the Hessian Matrix
\begin{equation}
 H_{n,m} \approx 2 \alpha_{n,m} = 2 \sum_{i=1}^N \frac{1}{\delta o_i^2} \frac{\partial t_i(k,\theta,(\lambda)^{JS}_{l,l'})}{\partial \lambda_n} \frac{\partial t_i(k,\theta,(\lambda)^{JS}_{l,l'})}{\partial \lambda_m}
 \label{eq:Hessian}
\end{equation}
This matrix is used to calculate the optimal change in parameters to
reduce the number of steps needed to minimize $\chi^2$. Once a minimum
has been found, inversion of the $\alpha$ matrix gives the usual
covariance matrix of the fitting parameters. An advantage of the
$\delta$-shell potential is that the derivatives in equation
(\ref{eq:Hessian}) can be computed analytically and simultaneously
with the corresponding observable. This approach greatly reduces the
numerical effort needed to minimize $\chi^2$. Moreover, it also
sidesteps the large inaccuracies trigered by a determination of the
covariance matrix using a numerical evaluation of second derivatives
and crossed derivatives at the minimum for a function with 46
parameters (we easily found non positive covariance matrices in this
way). Details of the numerical algorithm can be found in
\cite{press1992numerical}.

\section{Selection of data}
\label{sec:DataSelection}

\begin{figure*}[hpt]
\begin{center}
\epsfig{figure=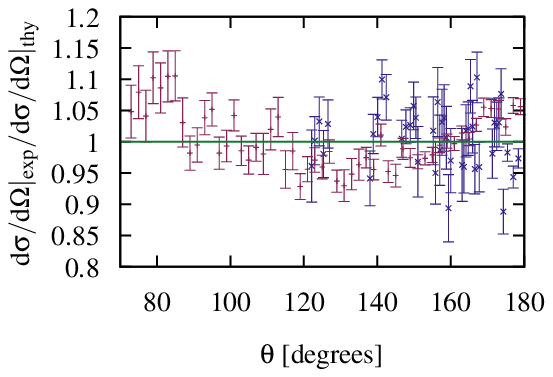,width=0.49\linewidth}  
\epsfig{figure=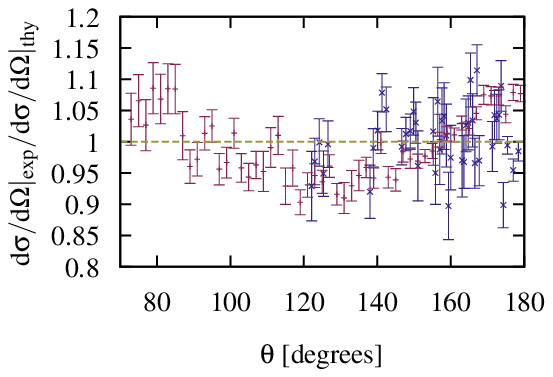,width=0.49\linewidth} \\
\vskip1ex
\epsfig{figure=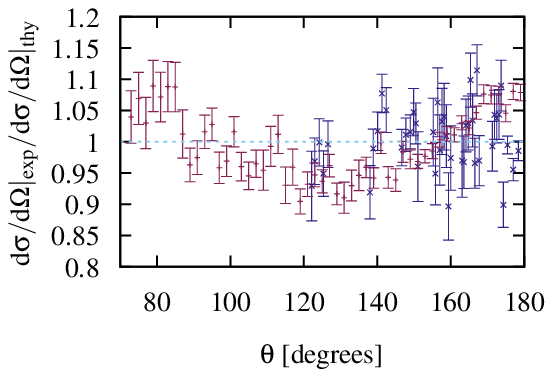,width=0.49\linewidth}
\epsfig{figure=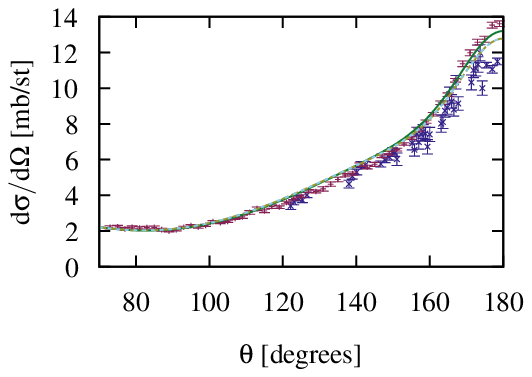,width=0.49\linewidth}
\end{center}
\caption{(Color online) np differential cross section at 162 MeV. The
top-left, top-right and bottom-left pannels compare the
experimental data sets of BO78\cite{PhysRevLett.41.1200} (blue crosses with error bars) and
RA98\cite{Rahm:1998jt} (red lines with error bars) to a fit to the complete
database (green solid line), to the initial and final consistent
databases (yellow dashed and light blue dotted lines) respectively as
discussed in the text. Every data point is scaled to the corresponding
fit. The bottom-right panel compares the unscaled data with the three
fits.}
\label{Fig:DSGat162}
\end{figure*}

Our database contains a total of $2972$ pp and $4737$ np published
scattering data up to the year 2013 with $E_{\rm LAB} \leq
350$MeV. Unfortunately there seems to be mutually incompatible data,
most likely due to under and overestimations of statistical and
systematic uncertainties from the experimental side. A clear example
of this inconsistencies, at backward angles in this case, are the two
data sets of np diferential cross section at 162 MeV
\cite{PhysRevLett.41.1200,Rahm:1998jt} plotted in the bottom right
panel of Fig.~\ref{Fig:DSGat162} for illustration purposes. To deal
with these inconsistencies we improve the $3\sigma$ criterion
introduced in this context by the Nijmegen group in their 1993 partial
wave analysis~\cite{Stoks:1993tb} which became an essential tenet of
their success and the subsequent high quality fits following
thereafter~\cite{Stoks:1994wp,Wiringa:1994wb,Machleidt:2000ge,Gross:2008ps}.
This criterion discards mutually incompatible data, but has the
unwanted side effect of eventually preventing a fraction of the data
to contribute positively to the final fit. This is so because no
distinction is made between mutually incompatible data sets in similar
kinematical conditions and which of them, if any, are actually
incompatible with the remaining data in different kinematical
conditions as encoded in the phenomenological parameterization which
intertwines all kinematical regions below pion production threshold.
We propose below an extended self-consistent $3\sigma$ criterion which
actually differenciates both situations.

Firstly, let us explain the traditional $3\sigma$ criterion used so
far in the literature claiming a final $\chi^2/{\rm d.o.f.} \lesssim 1
$. For a set of $n$ measurements with a Gaussian distribution, the
quantity $z \equiv \chi^2/n$ will satisfy the normalized probability
distribution
\begin{equation}
 \mathcal{P}_n(z) = \frac{n(n z/2)^{\frac{1}{2}-1}}{2 \Gamma(n/2)} e^{-nz/2}.
\end{equation}
According to the $3\sigma$ criterion a data set is considered
inconsistent with the rest of the database (More specifically, a
phenomenological model representing such database is meant in
practice), if $z$ has a probability smaller than a $0.27\%$. In most
cases a data set will have a highly improbable $z$ value if the
statistical errors are either underestimated ($z$ will be very high)
or overestimated ($z$ will be very low), then for every $n$ there is a
interval between $z_{\rm min}$ and $z_{\rm max}$ of allowed values of
$z$. Such endpoints are defined by
\begin{eqnarray}
  0.0027 &=& \int_0^{z_{\rm min}(n)} \mathcal{P}_n(z) dz = 1-\frac{\Gamma(n/2,nz_{\rm min}/2)}{\Gamma(n/2)}, \nonumber \\
  0.0027 &=& \int_{z_{\rm max}(n)}^\infty \mathcal{P}_n(z) dz = \frac{\Gamma(n/2,nz_{\rm max}/2)}{\Gamma(n/2)}.
\end{eqnarray}
The decision to float data with $\chi^2_{\rm sys} > 9 $ is a consequence
of applying the $3\sigma$ criterion with $n=1$ to the normalization
data.

Our selection process aims to obtain the largest possible database
that contains only consistent data with each other. To be able to
compare a single data set with the rest it is necessary to have a
model describing all of the available data as accurately as
possible. Therefore, we start by fitting our $\delta$-shell potential
to the complete database with $N = 2972|_{pp,\rm exp} + 159|_{pp,\rm
  norm} + 4737|_{np,\rm exp} + 259|_{np,\rm norm} = 3131|_{pp} +
4993|_{np}$ and obtain $\chi^2 = 3543.74|_{pp} + 8390.27|_{np}$ wich
yields $\chi^2/{\rm d.o.f.} = 1.48$.  This larger than one value was
expected since we know that mutually incompatible experimental data
are present. A comparison of the two np differential cross section
experiments at $162$MeV with this initial fit is shown on the top left
pannel in Fig.~\ref{Fig:DSGat162}. Note that at angles where the error
bars do not overlap, the model gives an ``in the middle'' solution
where the sum of both contributions to $\chi^2$ is minimized. Now
every data set can be tested using the $3\sigma$ criterion and
compared with the rest of the database via the first fit (in this case
both BO78\cite{PhysRevLett.41.1200} and RA98\cite{Rahm:1998jt} have $z
> z_{\rm max}$). Every data set failing to satisfy $z_{\rm min} \leq z
\leq z_{\rm max}$ is excluded and the remaining $N = 3008|_{pp} +
3438|_{np}$ data make what we call the initial and mutually consistent
database. By construction, this is a very close approximation to the
minimal mutually consistent data base.  A second fit is then
performed, this time to these initial and mutually compatible data
only, and $\chi^2 = 3061.97|_{pp} + 3634.34$ is obtained, which yields
$\chi^2/{\rm d.o.f.} = 1.05$. At this point the standard $3\sigma$
criterion stops. However, looking at the top right pannel of
Fig.~\ref{Fig:DSGat162} where the same experimental data are compared
to the second fit, one can notice that the theoretical model is now
closer to the RA98~\cite{Rahm:1998jt} values even though the latter
played no role in the determination of the fitting parameters. It is
fair then to ask if the discarded RA98\cite{Rahm:1998jt} data, or any
other of the initially rejected sets, is compatible with the initial
and mutually consistent data base. 

To analyze this point, we apply anew the $3 \sigma$ criterion to all
of the data sets using the second fit.  We find that this time $z_{\rm
  RA98} > z_{\rm max}$, while $z_{\rm BO78} < z_{\rm max}$ instead;
this initially discarded BO78\cite{PhysRevLett.41.1200} data set is
now recovered along with a total of 269 data and the parameters can be
refitted again. This particular example shows the potential good
features of the Gross and Stadler proposal~\cite{Gross:2008ps}.

Therefore, we apply this improved $3\sigma$ criterion systematically
to the full data base in a self consistent manner.  Namely, the
process can be repeated iteratively until no more data is recovered or
rejected. The bulk of recovered data is obtained the second time the
$3\sigma$ criterion is applied, for succesive steps only one or two
data sets are recovered or rejected. These statistical fluctuations
can be regarded as a marginal effect provided the range of variation
in the fitting parameters is substantially smaller than their final
quoted uncertainty. Our final fit meets this requirement.

A final and mutually consistent database with $N = 2996|_{pp} +
3717|_{np}$ data is obtained and the last re-fitting of the parameters
is carried out, yielding $\chi^2 = 3051.64|_{pp} + 3958.08|_{np}$
while the value of $\chi^2/{\rm d.o.f.} = 1.05$ is conserved. Finally
the bottom left pannel in Fig.~\ref{Fig:DSGat162} compares both
experiments with the last fit, but the differences with the top right
pannel are very small since the fitting parameters for the second fit
turn out to be very similar to the final
ones. Tables~\ref{tab:ppdatakeep} and~\ref{tab:npdatakeep} list all
the pp and np data included in the final and mutually consistent data
base to which the parameters $(\lambda)^{J,S}_{l,l'}$ are fitted.
Tables~\ref{tab:Rejectedpp} and~\ref{tab:Rejectednp} show the pp and
np rejected data.

We note that our final database includes {\it both} pp and
np. However, if we restrict to the np channel as done in
Ref.~\cite{Gross:2008ps} we find that those data close to the boundary
of {\it their} acceptance/rejection interval are also close to the
boundary of {\it our} acceptance/rejection interval, as the
corresponding data set chi square, $\chi^2_t$, are rather similar. The
inclusion or rejection in our case is supported by the pp observables.

\begin{longtable*}{@{\extracolsep{\fill}}l *{3}{c} *{2}{r} D{.}{.}{2.2} *{2}{D{.}{.}{1.3}} *{2}{d}}
\caption{\label{tab:ppdatakeep} pp scattering data sets used in the
  fit. The first column gives the energy, or energy range, of the
  experiment in MeV, the notation for the reference and the observable
  type is adopted from the SAID group \cite{SAID_NN}. $n_1$ represents
  the number of total measurements in the data set, while $n_2$ is the
  final number of measurements which can be different from $n_1$ due
  to rejection of outliers an the inclusion of normalization data. See
  alo main text.}
\\
\hline
\hline\noalign{\smallskip}
\footnotesize
  $E_{\rm LAB}$ &  Ref.       & Type &   $\theta$    &$n_1$&$n_2$& \multicolumn{1}{c}{\mbox{Sys}} & \multicolumn{1}{c}{$\chi^2_{\rm sys}$}&\multicolumn{1}{c}{$Z$} & \multicolumn{1}{c}{$\chi^2_t$} & \multicolumn{1}{c}{$\chi^2_t/n_2$}  \\ 
\hline\noalign{\smallskip}
\endfirsthead
\multicolumn{11}{c}{ \tablename\ \thetable{}. (Continued)} \\
\hline
\hline\noalign{\smallskip}
\footnotesize
  $E_{\rm LAB}$ &  Ref.       & Type &   $\theta$    &$n_1$&$n_2$& \multicolumn{1}{c}{\mbox{Sys}} & \multicolumn{1}{c}{$\chi^2_{\rm sys}$}&\multicolumn{1}{c}{$Z$} & \multicolumn{1}{c}{$\chi^2_t$} & \multicolumn{1}{c}{$\chi^2_t/n_2$}  \\ 
\hline\noalign{\smallskip}
\endhead
\hline
\endfoot
\noalign{\smallskip}\hline
\hline
\endlastfoot
\footnotesize
\input{keptpp}
\end{longtable*}

\begin{longtable*}{@{\extracolsep{\fill}}l *{3}{c} *{2}{r} *{5}{d}}
\caption{\label{tab:npdatakeep} Same as Table~\ref{tab:ppdatakeep} for np scattering data.} \\
\hline
\hline\noalign{\smallskip}
\footnotesize
  $E_{\rm LAB}$ &  Ref.       & Type &   $\theta$    &$n_1$&$n_2$& \multicolumn{1}{c}{\mbox{Sys}} & \multicolumn{1}{c}{$\chi^2_{\rm sys}$}&\multicolumn{1}{c}{$Z$} & \multicolumn{1}{c}{$\chi^2_t$} & \multicolumn{1}{c}{$\chi^2_t/n_2$}  \\ 
\hline\noalign{\smallskip}
\endfirsthead
\multicolumn{11}{c}{ \tablename\ \thetable{}. (Continued)} \\
\hline
\hline\noalign{\smallskip}
\footnotesize
  $E_{\rm LAB}$ &  Ref.       & Type &   $\theta$    &$n_1$&$n_2$& \multicolumn{1}{c}{\mbox{Sys}} & \multicolumn{1}{c}{$\chi^2_{\rm sys}$}&\multicolumn{1}{c}{$Z$} & \multicolumn{1}{c}{$\chi^2_t$} & \multicolumn{1}{c}{$\chi^2_t/n_2$}  \\ 
\hline\noalign{\smallskip}
\endhead
\hline
\endfoot
\noalign{\smallskip}\hline
\hline
\endlastfoot
\footnotesize
\input{keptnp}
\end{longtable*}

\begin{table*}
 \footnotesize
 \caption{\label{tab:Rejectedpp} Same as Table~\ref{tab:ppdatakeep} but for rejected pp data.}
 \begin{ruledtabular}
  \begin{tabular*}{\textwidth}{@{\extracolsep{\fill}}l *{3}{c} *{2}{r} D{.}{.}{2.2} *{2}{D{.}{.}{2.3}} D{.}{.}{1.6} D{.}{.}{1.6}}
  $E_{\rm LAB}$ &  Ref.       & Type &   $\theta$    &$n_1$&$n_2$& \multicolumn{1}{c}{\mbox{Sys}} & \multicolumn{1}{c}{$\chi^2_{\rm sys}$}&\multicolumn{1}{c}{$Z$} & \multicolumn{1}{c}{$\chi^2_t$} & \multicolumn{1}{c}{$\chi^2_t/n_2$}  \\ 
  \hline 
  0.3 -   0.4 & DO97\cite{Dombrowski:1997ja} & DSG  &      90.0     &  14 &  15 &  0.1\% & 4.083            & 0.998        &  318.20 & 21.21 \\
  0.4         & TH78\cite{Thomann1978457}    & DSG  &  58.4 -  95.0 &   4 &   4 & \multicolumn{2}{c}{Float} & 0.998        &   18.45 &  4.61 \\
  0.4         & TH78\cite{Thomann1978457}    & DSG  &  24.0 - 110.0 &  22 &  23 &  0.2\% & 0.016            & 1.000        &   62.88 &  2.73 \\
 14.2         & KI60\cite{JPSJ.15.749}       & DSG  &  18.1 - 114.2 &  17 &  18 & 10.0\% & 0.008            & 0.991        &    1.39 &  0.08 \\
 20.2         & CA67\cite{Catillon196793}    & P    &  34.8 -  90.3 &   8 &   9 & 11.8\% & 0.019            & 0.984        &    1.20 &  0.13 \\
 46.9         & GA71\cite{Garreta1971492}    & AXX  &      90.0     &   1 &   1 & \multicolumn{3}{c}{No systematic error}  &   15.73 & 15.73 \\
 66.0         & PA58\cite{Palmieri1958299}   & DSG  &  25.5 -  71.0 &  10 &  10 & \multicolumn{2}{c}{Float} & 1.037        &    1.78 &  0.18 \\
\end{tabular*}
 \end{ruledtabular}
\end{table*}

\begin{table*}
 \footnotesize
 \caption{\label{tab:Rejectednp} Same as Table~\ref{tab:ppdatakeep}
   but for rejected np data.}
 \begin{ruledtabular}
  \begin{tabular*}{\textwidth}{@{\extracolsep{\fill}}l *{3}{c} *{2}{r} D{.}{.}{2.2} *{2}{D{.}{.}{1.3}} D{.}{.}{3.2} D{.}{.}{2.2} }
  $E_{\rm LAB}$ &  Ref.       & Type &   $\theta$    &$n_1$&$n_2$& \multicolumn{1}{c}{\mbox{Sys}} & \multicolumn{1}{c}{$\chi^2_{\rm sys}$}&\multicolumn{1}{c}{$Z$} & \multicolumn{1}{c}{$\chi^2_t$} & \multicolumn{1}{c}{$\chi^2_t/n_2$}  \\ 
  \hline 
  0.5 -   2.0 & PO82\cite{Poenitz1982224}       & SGT  &       -       &   3 &   3 & \multicolumn{3}{c}{No systematic error}  &   30.24 & 10.08 \\
  1.0 -   2.5 & FI54\cite{PhysRev.94.389}       & SGT  &       -       &   2 &   2 & \multicolumn{3}{c}{No systematic error}  &   14.51 &  7.25 \\
  2.5         & DV71\cite{PhysRevC.4.1061}      & SGT  &       -       &   1 &   1 & \multicolumn{3}{c}{No systematic error}  &   14.04 & 14.04 \\
 12.0         & BR08\cite{Braun2008161}         & P    &  32.6 - 143.8 &  15 &  16 &  1.5\% & 5.271            & 0.966        &   41.63 &  2.60 \\
 14.1         & SU67\cite{Suhami1967173}        & DSG  &  11.9 -  92.8 &  16 &  16 & \multicolumn{2}{c}{Float} & 0.999        &    3.19 &  0.20 \\
 31.5         & BE97\cite{Benck1997220}         & DSG  &  40.0 - 140.0 &   6 &   7 &  5.0\% & 0.014            & 1.006        &   57.54 &  8.22 \\
 37.5         & BE97\cite{Benck1997220}         & DSG  &  40.0 - 140.0 &   6 &   7 &  5.0\% & 0.105            & 0.984        &   25.82 &  3.69 \\
\end{tabular*}
 \end{ruledtabular}
\end{table*}

\section{Results}
\label{sec:Results}

With a consistent database it is posible to properly estimate
statistical errors in the potential parameters. In
Ref.~\cite{Perez:2013mwa} we give the corresponding partial wave
independent fitting parameters which, as explained above, proved more
convenient as primary quantities to carry out the fit due to smaller
statistical correlations. Here, table \ref{tab:Fits} shows the
strength coefficients $V_{i,n}$ and their statistical uncertainties
propagated from the experimental data via the usual covariance matrix
and applying Appendix~\ref{sec:transform} to the results of
Ref.~\cite{Perez:2013mwa}. With these parameters and the covariance
matrix it is possible to also estimate and propagate statistical error
bars for calculations made with the $\delta$-shell potential. For
example tables~\ref{tab:ppIsovectorPS}, \ref{tab:npIsovectorPS} and~,
\ref{tab:npIsoscalarPS} show pp isovector, np isovector and np isoscalar
phaseshifts respectively with statistical errors extracted from
experimental data for a few partial waves at different kinetic
laboratory frame energy. Comparing our results with the Tables IV and
V in~\cite{Stoks:1993tb} we find that for low angular momentum partial waves
our estimates for the errors tend to be smaller than the ones obtained
by the Nijmegen group in their 1993 PWA, while for higher values of
$l$ the Nijmegen errors are smaller.

\begingroup
\squeezetable
\begin{table*}[htb]
% \footnotesize
 \caption{\label{tab:Fits} Delta-shell potential parameters $V_{i,n} $
   (in ${\rm fm}^{-1}$) with their errors for all operators. This
   parameters are {\it not} the fitting parameters and are obtained
   from the 46 {\it independent} partial waves
   $(\lambda_i)^{JS}_{l,l'}$ (see Appendix \ref{sec:transform}). We
   take $N=5$ equidistant points with $\Delta r = 0.6$fm. Rows marked
   with $^*$ indicates that the corresponding strengths coefficients
   are not independent or zero. In the first line we provide the central
   component of the delta shells corresponding to the EM effects below
   $r_c = 3 {\rm fm}$. These parameters remain fixed within the
   fitting process.}
  \begin{ruledtabular}      
	\begin{tabular*}{\columnwidth}{@{\extracolsep{\fill}}c *{4}{D{.}{.}{2.7}} *{1}{D{.}{.}{2.9}} }
 Operator  & \multicolumn{1}{c}{$V_{1,x}$} & \multicolumn{1}{c}{$V_{2,x}$} & \multicolumn{1}{c}{$V_{3,x}$} & \multicolumn{1}{c}{$V_{4,x}$} & \multicolumn{1}{c}{$V_{5,x}$}  \\
       & \multicolumn{1}{c}{$r_1=0.6$fm} & \multicolumn{1}{c}{$r_2=1.2$fm} & \multicolumn{1}{c}{$r_3=1.8$fm} & \multicolumn{1}{c}{$r_4=2.4$fm} & \multicolumn{1}{c}{$r_5=3.0$fm}  \\
            \hline
$V_C[{\rm pp}]_{\rm EM}$  & 0.0073308  &   0.0063680  &   0.0033378  &   0.0036882  &   0.0009250 \\ 
%\hline 
\noalign{\smallskip}
 $c$               &   0.195(3)  &   -0.059(2)   &  -0.0125(4)  &  -0.0059(2)  &  -0.00215(7)  \\
 $\tau$            &  -0.085(3)  &   -0.021(1)   &  -0.0035(2)  &   0.0025(1)  &  -0.00072(2)  \\
 $\sigma$          &  -0.045(2)  &    0.0029(8)  &   0.0120(3)  &  -0.00405(9) &   0.00122(3)  \\
 $\sigma\tau$      &  -0.0649(9) &    0.0563(6)  &   0.0033(1)  &   0.00292(7) &   0.00041(1)  \\
 $t$               &   0.0       &   -0.026(2)   &   0.0067(5)  &  -0.0026(2)  &   0.00047(8)  \\
 $t \tau$          &   0.0       &     0.0591(6) &   0.0168(4)  &   0.0083(2)  &   0.00112(7)  \\
 $ls$              &   0.0       &    -0.198(2)  &   0.0169(4)  &  -0.0053(2)  &   0.00080(5)  \\
 $ls \tau$         &   0.0       &    -0.0837(8) &   0.0112(3)  &  -0.0029(1)  &   0.00041(4)  \\
 $l2$              &  -0.0325(5) &    0.075(2)   &  -0.0099(3)  &   0.0030(1)  &  -0.00046(3)  \\
 $l2 \tau$         &   0.0141(5) &    0.0156(7)  &  -0.0041(2)  &   0.00084(6) &  -0.00007(1)  \\
 $l2 \sigma$       &   0.0075(4) &    0.0207(7)  &  -0.0032(1)  &   0.00094(5) &  -0.00020(1)  \\
 $l2 \sigma\tau$   &   0.0108(2) &   -0.0009(3)  &  -0.00091(8) &   0.00034(3) &  -0.000039(6) \\
 $ls2$             &   0.0       &   -0.066(3)   &   0.0134(4)  &  -0.0037(2)  &   0.00068(5)  \\
 $ls2 \tau$        &   0.0       &   -0.007(1)   &   0.0047(3)  &  -0.00134(9) &   0.00013(3)  \\
 $T$               &   0.0022(9) &    0.0007(3)  &  -0.0005(1)  &   0.0        &   0.00005(2)  \\
 $\sigma T^*$      &  -0.0022(9) &   -0.0007(3)  &   0.0005(1)  &   0.0        &  -0.00005(2)  \\
 $t T^*$           &   0.0       &    0.0        &   0.0        &   0.0        &   0.0         \\
 $\tau z^*$        &   0.0       &    0.0        &   0.0        &   0.0        &   0.0         \\
 $\sigma\tau z^*$  &   0.0       &    0.0        &   0.0        &   0.0        &   0.0         \\
 $l2 T^*$          &  -0.0004(2) &   -0.00012(4) &   0.00009(2) &   0.0        &  -0.000008(3) \\
 $l2 \sigma T^*$   &   0.0004(2) &    0.00012(4) &  -0.00009(2) &   0.0        &   0.000008(3) \\
	\end{tabular*}
 \end{ruledtabular}	
\end{table*}
\endgroup

\begin{table*}
 \footnotesize
 \caption{\label{tab:ppIsovectorPS} pp isovector phaseshifts.}
 \begin{ruledtabular}
 \begin{tabular*}{\textwidth}{@{\extracolsep{\fill}} r *{12}{D{.}{.}{3.3}}}
$E_{\rm LAB}$&\multicolumn{1}{c}{$^1S_0$}&\multicolumn{1}{c}{$^1D_2$}&\multicolumn{1}{c}{$^1G_4$}&\multicolumn{1}{c}{$^3P_0$}&\multicolumn{1}{c}{$^3P_1$}&\multicolumn{1}{c}{$^3F_3$}&\multicolumn{1}{c}{$^3P_2$}&\multicolumn{1}{c}{$\epsilon_2$}&\multicolumn{1}{c}{$^3F_2$}&\multicolumn{1}{c}{$^3F_4$}&\multicolumn{1}{c}{$\epsilon_4$}&\multicolumn{1}{c}{$^3H_4$}\\ 
  \hline 
  1 &    32.651 &     0.001 &     0.000 &     0.133 &    -0.080 &    -0.000 &     0.014 &    -0.001 &     0.000 &     0.000 &    -0.000 &     0.000\\
    & \pm 0.003 & \pm 0.000 & \pm 0.000 & \pm 0.000 & \pm 0.000 & \pm 0.000 & \pm 0.000 & \pm 0.000 & \pm 0.000 & \pm 0.000 & \pm 0.000 & \pm 0.000\\
  5 &    54.841 &     0.042 &     0.000 &     1.582 &    -0.892 &    -0.004 &     0.213 &    -0.052 &     0.002 &     0.000 &    -0.000 &     0.000\\
    & \pm 0.006 & \pm 0.000 & \pm 0.000 & \pm 0.003 & \pm 0.001 & \pm 0.000 & \pm 0.001 & \pm 0.000 & \pm 0.000 & \pm 0.000 & \pm 0.000 & \pm 0.000\\
 10 &    55.256 &     0.164 &     0.003 &     3.737 &    -2.037 &    -0.031 &     0.648 &    -0.201 &     0.013 &     0.001 &    -0.004 &     0.000\\
    & \pm 0.010 & \pm 0.000 & \pm 0.000 & \pm 0.008 & \pm 0.002 & \pm 0.000 & \pm 0.002 & \pm 0.000 & \pm 0.000 & \pm 0.000 & \pm 0.000 & \pm 0.000\\
 25 &    48.789 &     0.694 &     0.040 &     8.607 &    -4.862 &    -0.230 &     2.489 &    -0.811 &     0.105 &     0.019 &    -0.049 &     0.004\\
    & \pm 0.013 & \pm 0.001 & \pm 0.000 & \pm 0.020 & \pm 0.006 & \pm 0.000 & \pm 0.005 & \pm 0.001 & \pm 0.000 & \pm 0.000 & \pm 0.000 & \pm 0.000\\
 50 &    39.185 &     1.709 &     0.152 &    11.524 &    -8.208 &    -0.682 &     5.854 &    -1.708 &     0.341 &     0.102 &    -0.197 &     0.026\\
    & \pm 0.017 & \pm 0.004 & \pm 0.000 & \pm 0.032 & \pm 0.012 & \pm 0.002 & \pm 0.009 & \pm 0.004 & \pm 0.001 & \pm 0.001 & \pm 0.000 & \pm 0.000\\
100 &    25.444 &     3.781 &     0.418 &     9.497 &   -13.259 &    -1.473 &    10.980 &    -2.643 &     0.832 &     0.461 &    -0.554 &     0.110\\
    & \pm 0.033 & \pm 0.008 & \pm 0.002 & \pm 0.056 & \pm 0.019 & \pm 0.009 & \pm 0.018 & \pm 0.008 & \pm 0.008 & \pm 0.005 & \pm 0.001 & \pm 0.000\\
150 &    15.254 &     5.618 &     0.703 &     4.816 &   -17.620 &    -2.061 &    14.008 &    -2.913 &     1.192 &     1.013 &    -0.877 &     0.220\\
    & \pm 0.046 & \pm 0.014 & \pm 0.006 & \pm 0.066 & \pm 0.025 & \pm 0.019 & \pm 0.019 & \pm 0.012 & \pm 0.017 & \pm 0.009 & \pm 0.003 & \pm 0.002\\
200 &     7.041 &     7.192 &     1.006 &    -0.227 &   -21.506 &    -2.505 &    15.784 &    -2.901 &     1.327 &     1.634 &    -1.129 &     0.341\\
    & \pm 0.057 & \pm 0.021 & \pm 0.012 & \pm 0.062 & \pm 0.036 & \pm 0.033 & \pm 0.025 & \pm 0.016 & \pm 0.022 & \pm 0.016 & \pm 0.005 & \pm 0.006\\
250 &     0.318 &     8.538 &     1.307 &    -5.121 &   -24.903 &    -2.718 &    16.715 &    -2.703 &     1.223 &     2.200 &    -1.308 &     0.464\\
    & \pm 0.074 & \pm 0.024 & \pm 0.017 & \pm 0.063 & \pm 0.051 & \pm 0.046 & \pm 0.032 & \pm 0.022 & \pm 0.027 & \pm 0.022 & \pm 0.006 & \pm 0.012\\
300 &    -5.098 &     9.577 &     1.586 &    -9.663 &   -27.811 &    -2.547 &    16.951 &    -2.312 &     0.921 &     2.644 &    -1.434 &     0.572\\
    & \pm 0.100 & \pm 0.031 & \pm 0.019 & \pm 0.089 & \pm 0.064 & \pm 0.054 & \pm 0.033 & \pm 0.030 & \pm 0.039 & \pm 0.027 & \pm 0.007 & \pm 0.019\\
350 &    -9.342 &    10.183 &     1.842 &   -13.677 &   -30.221 &    -1.922 &    16.600 &    -1.730 &     0.461 &     2.970 &    -1.541 &     0.650\\
    & \pm 0.134 & \pm 0.052 & \pm 0.026 & \pm 0.141 & \pm 0.076 & \pm 0.068 & \pm 0.030 & \pm 0.039 & \pm 0.055 & \pm 0.042 & \pm 0.010 & \pm 0.026\\
 \end{tabular*}
 \end{ruledtabular}
\end{table*}

\begin{table*}
 \footnotesize
 \caption{\label{tab:npIsovectorPS}np isovector phaseshifts.}
 \begin{ruledtabular}
  \begin{tabular*}{\textwidth}{@{\extracolsep{\fill}} r *{12}{D{.}{.}{3.3}}}
$E_{\rm LAB}$&\multicolumn{1}{c}{$^1S_0$}&\multicolumn{1}{c}{$^1D_2$}&\multicolumn{1}{c}{$^1G_4$}&\multicolumn{1}{c}{$^3P_0$}&\multicolumn{1}{c}{$^3P_1$}&\multicolumn{1}{c}{$^3F_3$}&\multicolumn{1}{c}{$^3P_2$}&\multicolumn{1}{c}{$\epsilon_2$}&\multicolumn{1}{c}{$^3F_2$}&\multicolumn{1}{c}{$^3F_4$}&\multicolumn{1}{c}{$\epsilon_4$}&\multicolumn{1}{c}{$^3H_4$}\\ 
  \hline 
  1 &    62.077 &     0.001 &     0.000 &     0.181 &    -0.107 &    -0.000 &     0.022 &    -0.001 &     0.000 &     0.000 &    -0.000 &     0.000\\
    & \pm 0.018 & \pm 0.000 & \pm 0.000 & \pm 0.000 & \pm 0.000 & \pm 0.000 & \pm 0.000 & \pm 0.000 & \pm 0.000 & \pm 0.000 & \pm 0.000 & \pm 0.000\\
  5 &    63.660 &     0.041 &     0.000 &     1.657 &    -0.933 &    -0.004 &     0.258 &    -0.048 &     0.002 &     0.000 &    -0.000 &     0.000\\
    & \pm 0.044 & \pm 0.000 & \pm 0.000 & \pm 0.003 & \pm 0.001 & \pm 0.000 & \pm 0.001 & \pm 0.000 & \pm 0.000 & \pm 0.000 & \pm 0.000 & \pm 0.000\\
 10 &    60.020 &     0.155 &     0.002 &     3.754 &    -2.057 &    -0.026 &     0.727 &    -0.184 &     0.011 &     0.001 &    -0.003 &     0.000\\
    & \pm 0.063 & \pm 0.000 & \pm 0.000 & \pm 0.008 & \pm 0.002 & \pm 0.000 & \pm 0.002 & \pm 0.000 & \pm 0.000 & \pm 0.000 & \pm 0.000 & \pm 0.000\\
 25 &    51.071 &     0.674 &     0.032 &     8.479 &    -4.877 &    -0.198 &     2.627 &    -0.763 &     0.091 &     0.016 &    -0.039 &     0.003\\
    & \pm 0.099 & \pm 0.001 & \pm 0.000 & \pm 0.021 & \pm 0.006 & \pm 0.000 & \pm 0.005 & \pm 0.001 & \pm 0.000 & \pm 0.000 & \pm 0.000 & \pm 0.000\\
 50 &    40.885 &     1.702 &     0.132 &    11.300 &    -8.318 &    -0.611 &     6.011 &    -1.661 &     0.308 &     0.092 &    -0.170 &     0.021\\
    & \pm 0.147 & \pm 0.004 & \pm 0.001 & \pm 0.033 & \pm 0.013 & \pm 0.002 & \pm 0.009 & \pm 0.004 & \pm 0.002 & \pm 0.001 & \pm 0.000 & \pm 0.000\\
100 &    27.210 &     3.775 &     0.370 &     9.140 &   -13.571 &    -1.381 &    11.062 &    -2.665 &     0.775 &     0.437 &    -0.506 &     0.093\\
    & \pm 0.230 & \pm 0.008 & \pm 0.009 & \pm 0.057 & \pm 0.019 & \pm 0.009 & \pm 0.018 & \pm 0.008 & \pm 0.009 & \pm 0.005 & \pm 0.001 & \pm 0.000\\
150 &    17.100 &     5.571 &     0.606 &     4.336 &   -18.034 &    -2.004 &    14.004 &    -2.982 &     1.110 &     0.971 &    -0.829 &     0.194\\
    & \pm 0.266 & \pm 0.014 & \pm 0.030 & \pm 0.066 & \pm 0.025 & \pm 0.020 & \pm 0.020 & \pm 0.012 & \pm 0.017 & \pm 0.010 & \pm 0.003 & \pm 0.002\\
200 &     8.830 &     7.106 &     0.849 &    -0.796 &   -21.958 &    -2.509 &    15.697 &    -2.971 &     1.212 &     1.563 &    -1.095 &     0.309\\
    & \pm 0.267 & \pm 0.022 & \pm 0.059 & \pm 0.062 & \pm 0.037 & \pm 0.033 & \pm 0.026 & \pm 0.017 & \pm 0.022 & \pm 0.016 & \pm 0.005 & \pm 0.006\\
250 &     1.971 &     8.424 &     1.119 &    -5.749 &   -25.375 &    -2.781 &    16.542 &    -2.748 &     1.074 &     2.096 &    -1.295 &     0.423\\
    & \pm 0.281 & \pm 0.024 & \pm 0.081 & \pm 0.064 & \pm 0.051 & \pm 0.046 & \pm 0.032 & \pm 0.022 & \pm 0.028 & \pm 0.022 & \pm 0.006 & \pm 0.012\\
300 &    -3.618 &     9.432 &     1.438 &   -10.325 &   -28.308 &    -2.649 &    16.697 &    -2.328 &     0.744 &     2.513 &    -1.444 &     0.518\\
    & \pm 0.344 & \pm 0.032 & \pm 0.085 & \pm 0.091 & \pm 0.065 & \pm 0.054 & \pm 0.033 & \pm 0.030 & \pm 0.040 & \pm 0.027 & \pm 0.007 & \pm 0.019\\
350 &    -8.044 &     9.994 &     1.829 &   -14.349 &   -30.747 &    -2.040 &    16.280 &    -1.727 &     0.260 &     2.822 &    -1.571 &     0.578\\
    & \pm 0.453 & \pm 0.054 & \pm 0.083 & \pm 0.145 & \pm 0.076 & \pm 0.069 & \pm 0.030 & \pm 0.040 & \pm 0.056 & \pm 0.043 & \pm 0.010 & \pm 0.026\\
 \end{tabular*}
 \end{ruledtabular}
\end{table*}

\begin{table*}
 \footnotesize
 \caption{\label{tab:npIsoscalarPS}np isoscalar phaseshifts.}
 \begin{ruledtabular}
 \begin{tabular*}{\textwidth}{@{\extracolsep{\fill}} r *{12}{D{.}{.}{3.3}}}
 $E_{\rm LAB}$&\multicolumn{1}{c}{$^1P_1$}&\multicolumn{1}{c}{$^1F_3$}&\multicolumn{1}{c}{$^3D_2$}&\multicolumn{1}{c}{$^3G_4$}&\multicolumn{1}{c}{$^3S_1$}&\multicolumn{1}{c}{$\epsilon_1$}&\multicolumn{1}{c}{$^3D_1$}&\multicolumn{1}{c}{$^3D_3$}&\multicolumn{1}{c}{$\epsilon_3$}&\multicolumn{1}{c}{$^3G_3$}\\ 
  \hline 
  1 &    -0.186 &    -0.000 &     0.006 &     0.000 &   147.647 &     0.104 &    -0.005 &     0.000 &     0.000 &    -0.000\\
    & \pm 0.000 & \pm 0.000 & \pm 0.000 & \pm 0.000 & \pm 0.010 & \pm 0.001 & \pm 0.000 & \pm 0.000 & \pm 0.000 & \pm 0.000\\
  5 &    -1.494 &    -0.010 &     0.219 &     0.001 &   117.954 &     0.657 &    -0.180 &     0.002 &     0.012 &    -0.000\\
    & \pm 0.004 & \pm 0.000 & \pm 0.000 & \pm 0.000 & \pm 0.022 & \pm 0.004 & \pm 0.000 & \pm 0.000 & \pm 0.000 & \pm 0.000\\
 10 &    -3.062 &    -0.064 &     0.847 &     0.012 &   102.292 &     1.125 &    -0.671 &     0.005 &     0.080 &    -0.003\\
    & \pm 0.010 & \pm 0.000 & \pm 0.001 & \pm 0.000 & \pm 0.031 & \pm 0.009 & \pm 0.002 & \pm 0.000 & \pm 0.000 & \pm 0.000\\
 25 &    -6.357 &    -0.421 &     3.729 &     0.170 &    80.136 &     1.734 &    -2.788 &     0.043 &     0.553 &    -0.053\\
    & \pm 0.033 & \pm 0.000 & \pm 0.008 & \pm 0.000 & \pm 0.044 & \pm 0.020 & \pm 0.009 & \pm 0.001 & \pm 0.000 & \pm 0.000\\
 50 &    -9.663 &    -1.142 &     9.057 &     0.727 &    62.160 &     2.057 &    -6.423 &     0.315 &     1.615 &    -0.265\\
    & \pm 0.069 & \pm 0.003 & \pm 0.027 & \pm 0.001 & \pm 0.053 & \pm 0.037 & \pm 0.026 & \pm 0.004 & \pm 0.003 & \pm 0.001\\
100 &   -14.207 &    -2.304 &    17.387 &     2.248 &    42.712 &     2.473 &   -12.198 &     1.444 &     3.498 &    -0.996\\
    & \pm 0.106 & \pm 0.022 & \pm 0.049 & \pm 0.010 & \pm 0.053 & \pm 0.068 & \pm 0.048 & \pm 0.016 & \pm 0.016 & \pm 0.009\\
150 &   -17.977 &    -3.178 &    21.962 &     3.892 &    30.392 &     2.938 &   -16.407 &     2.643 &     4.825 &    -1.919\\
    & \pm 0.119 & \pm 0.055 & \pm 0.066 & \pm 0.033 & \pm 0.058 & \pm 0.095 & \pm 0.058 & \pm 0.033 & \pm 0.028 & \pm 0.030\\
200 &   -21.235 &    -3.953 &    24.017 &     5.484 &    20.936 &     3.448 &   -19.682 &     3.485 &     5.749 &    -2.879\\
    & \pm 0.144 & \pm 0.086 & \pm 0.085 & \pm 0.057 & \pm 0.072 & \pm 0.112 & \pm 0.071 & \pm 0.051 & \pm 0.034 & \pm 0.053\\
250 &   -24.030 &    -4.672 &    24.846 &     6.885 &    12.945 &     4.036 &   -22.324 &     3.931 &     6.427 &    -3.799\\
    & \pm 0.185 & \pm 0.102 & \pm 0.096 & \pm 0.070 & \pm 0.087 & \pm 0.111 & \pm 0.080 & \pm 0.071 & \pm 0.043 & \pm 0.069\\
300 &   -26.418 &    -5.283 &    25.324 &     7.986 &     5.837 &     4.753 &   -24.369 &     4.099 &     6.938 &    -4.669\\
    & \pm 0.226 & \pm 0.104 & \pm 0.114 & \pm 0.078 & \pm 0.098 & \pm 0.113 & \pm 0.108 & \pm 0.095 & \pm 0.053 & \pm 0.084\\
350 &   -28.418 &    -5.704 &    25.950 &     8.729 &    -0.644 &     5.610 &   -25.699 &     4.121 &     7.294 &    -5.517\\
    & \pm 0.261 & \pm 0.125 & \pm 0.189 & \pm 0.104 & \pm 0.105 & \pm 0.169 & \pm 0.191 & \pm 0.121 & \pm 0.064 & \pm 0.118\\
 \end{tabular*}
 \end{ruledtabular}
\end{table*}

%phasehifts (figures and tables)

Deuteron wavefunctions, static properties and form factors can be
calculated with the parameters of the $^3S_1$-$^3D_1$ coupled channel
for the bound state and the errors have been propagated. The results
have already been shown in our previous work~\cite{Perez:2013mwa} and
will not be discussed further here.

\section{Conclusions and outlook}
\label{sec:Conclusions}

The main problem in analysing the existing np and pp scattering data
lies in the large number of experiments in different kinematical
regions measuring a variety of observables with different accuracies
and a largely heterogenous abundance in the $(E_{\rm LAB},\theta)$
plane. On the other hand, for a given LAB energy and scattering angle
just five complex quantities suffice to describe the scattering
amplitude. Therefore the measurement of ten different observables at
a given point in this plane would reconstruct the full amplitude up to a
global phase. In terms of a partial wave decomposition at a fixed
energy one needs about as many angles as partial waves are needed for
the scattering amplitude to converge. For the energies listed in the
table this gives a total number of points in the $(E_{\rm
  LAB},\theta)$ plane much smaller than the number of available data.
Unfortunatelly, the experiments were not designed from this point of
view, and a rather different non-homogeneous distribution is at our
disposal. In addition, there are mutually inconsistent experiments and
a proper selection of data must be carried out. We have explored with
success an interesting criterion suggested by Gross and
Stadler~\cite{Gross:2008ps} in their covariant spectator theory
analysis of np scattering data. This criterion improves with respect
to the more customary $3\sigma$-criterion applied since the 1993
benchmarking fit.

The best known way of smoothly interpolating energy values between the
experimentally measured ones is to assume a phenomenological
potential. Of course, there exist conditions regarding the analytical
behaviour of the scattering amplitude and its partial wave
contributions which become relevant at sufficiently low energies and
depend only on the long distance behaviour of the potential and for
the case of OPE corresponds to the appearance of a left hand cut at
$p_{\rm CM}^2= - m_\pi^2 /4$ in the partial wave amplitudes. In this
paper we have used a rather simple phenomenological potential which
incorporates indisputable physical effects in their certified domain
of applicability and a unknown contribution which summarizes our
ignorance of the intermediate and short range parts. Our analysis is
consistent with the venerable and CD-OPE contribution above a distance
of 3 fm. The simplicity of the potential should not be confused as
unrealistic, as by definition our ignorance can be parameterized
according to anyone's prejudices. We choose to implement the appealing
idea of coarse graining of the interaction, since the shortest
distance which can be probed below pion production threshold
corresponds to a NN relative de Broglie wavelength resolution of
$\Delta r \sim 1 /\sqrt{m_\pi M_N} \sim 0.6 {\rm fm}$.  This simple
scale consideration along with the role of the centrifugal barrier
allows to foresee the number of needed fitting parameters for the
intermediate range
part~\cite{NavarroPerez:2011fm,NavarroPerez2013138,Perez:2013mwa}
which turns out to be on the order of the actual number of parameters.

A successful fit to pp and np scattering data bellow pion production
threshold was achieved using a $\delta$-shell representation for the
short and intermadiate range part of the nucleon-nucleon interaction;
the long range part was described by charge dependent OPE and EM
interactions. 

A partial wave decomposition of the potential allowed to incorporate
charge dependence in the $^1S_0$ parameters while the rest of the
parameters are charge independent. A comprehensive and ready-to-use
review of the necessary theory to compute the full on-shell
electromagnetic scattering amplitude was made. This includes every
single EM contribution, as well as the energy dependent parameter
$\alpha'$ and, as already found out in the previous high quality
analysis, turns out to be crucial for the correct description of pp
and np scattering observables. The $3\sigma$ criterion was used
iteratively to obtain a fully consistent database with $N_{\rm data} =
6713$ including normalization data. The complete data selection
process described here allowed to recover a total of $300$ data that a
single application of the $3\sigma$ criterion would instead have
rejected. The fitting parameters, their uncertainties and correlations
were determined from the final and mutually consistent database. Thus,
errors in the potential can be propagated to any other quantities
obtained using the $\delta$-shell representation. Phase shifts were
extracted from the data and are presented here with the corresponding
statistical error bars.  With a consistent database it is also possible to
test the statistical significance of the inclusion of other potentials
where the intermediate range contributions are explicitly modelled.

\appendix
\section{Tranformations between partial wave and operator basis}
\label{sec:transform}

As mentioned in the main text we use an operator basis extending the
AV18 potentials in coordinate space. We use the following definitions 
\begin{eqnarray}
S_{12} &=& 3 {\bf \sigma}_{1} \cdot \hat {\bf r} {\bf \sigma}_{2} \cdot \hat {\bf r} - {\bf \sigma}_{1} \cdot  {\bf \sigma}_{2}  \\ 
T_{12} &=& 3\tau_{z1}\tau_{z2}- \tau_{1}\!\cdot\!\tau_{2}
\end{eqnarray}
Thus, the total potential in the operator basis reads 
\begin{equation}
       V=\sum_{n=1,21} V_{n}(r ) O^{n} \ .
\end{equation}
Here, the first fourteen operators are the same charge-independent ones
used in the Argonne $v_{14}$ potential and are given by
\begin{eqnarray}
   O^{n=1,14} &=& 1,  {\tau}_{1}\!\cdot\! {\tau}_{2},\,
        {\sigma}_{1}\!\cdot\! {\sigma}_{2},
       ( {\sigma}_{1}\!\cdot\! {\sigma}_{2})
       ( {\tau}_{1}\!\cdot\! {\tau}_{2}),\,  S_{12}, S_{12}( {\tau}_{1}\!\cdot\! {\tau}_{2}),\, \nonumber \\ 
    &&
    {\bf L}\!\cdot\!{\bf S}, {\bf L}\!\cdot\!{\bf S}
        ( {\tau}_{1}\!\cdot\! {\tau}_{2}), L^{2}, L^{2}( {\tau}_{1}\!\cdot\! {\tau}_{2}),\,
    L^{2}( {\sigma}_{1}\!\cdot\! {\sigma}_{2}), \nonumber\\
  & & 
    L^{2}( {\sigma}_{1}\!\cdot\! {\sigma}_{2})
         ( {\tau}_{1}\!\cdot\! {\tau}_{2}),\, 
    ({\bf L}\!\cdot\!{\bf S})^{2}, ({\bf L}\!\cdot\!{\bf S})^{2}
        ( {\tau}_{1}\!\cdot\! {\tau}_{2})\ . \nonumber \\ 
\end{eqnarray}
These fourteen components are denoted by the abbreviations $c$,
$\tau$, $\sigma$, $\sigma\tau$, $t$, $t\tau$, $ls$, $ls\tau$, $l2$,
$l2\tau$, $l2\sigma$, $l2\sigma\tau$, $ls2$, and $ls2\tau$. The remaining 
terms are
\begin{eqnarray}
   O^{n=15,21}  &=&  T_{12}, \,
        ( {\sigma}_{1}\!\cdot\! {\sigma}_{2})
       , T_{12}\, S_{12}T_{12},\, (\tau_{z1}+\tau_{z2})\ , \nonumber \\ 
&&( {\sigma}_{1}\!\cdot\! {\sigma}_{2}) (\tau_{z1}+\tau_{z2})\ ,
L^{2} T_{12} ,  L^{2} ( {\sigma}_{1}\!\cdot\! {\sigma}_{2}) T_{12} \, .  \nonumber \\ 
\end{eqnarray}
These terms are charge dependent and are labeled as $T$, $\sigma
T$,$tT$, $\sigma \tau z$, $l2T$ and $l2\sigma T$.

Since we use the low angular momentum partial wave strength
coefficients as fitting parameters we need a transformation to
construct the complete potential and extract the high angular momentum
partial wave parameters from the latter. The high angular momentum
partial waves up to $J=20$ are important to calculate the strong
on-shell scattering amplitude. The transformation from the operator
basis to partial waves is direct and is given by

\begin{widetext}
 \begin{align}
   \label{eq:PWOperators}
  (\lambda_i)_{l,l'}^{J,S} = & \ \mu_{\alpha \beta} \Delta r \left[1-(-1)^{l+S+T}\right] \left[\vphantom{\frac{\sqrt{J}\delta_{J,}}{J}} \delta_{l,l'} \left( \vphantom{\frac{1}{2}} V_{i,c} + (4T-3)V_{i,\tau} + (4S-3)V_{i,\sigma} + (4S-3)(4T-3)V_{i,\sigma\tau} \right. \right. \\ \nonumber 
 + & \  \left. \left.  l(l+1)\left[V_{i,l2} + (4T-3)V_{i,l2\tau} + (4S-3)V_{i,l2\sigma} + (4S-3)(4T-3)V_{i,l2\sigma\tau}  \right] \right. \right. \\ \nonumber 
 + & \ \left. \left. \frac{1}{2}\left[J(J+1) - l(l+1) - S(S+1) \right]\left[V_{i,ls} + (4T-3)V_{i,ls\tau} \right] 
 + \frac{1}{4}\left[J(J+1) - l(l+1) - S(S+1) \right]^2\left[V_{i,ls2} + (4T-3)V_{i,ls2\tau} \right] \right. \right. \\ \nonumber 
 + & \ \left. \left. \delta_{1,T} \left[3\tau_{zi}\tau_{zj}-(4T-3) \right] \left[V_{i,T} + (4S-3)V_{i,\sigma T} + l(l+1)\left(V_{i,l2T} + (4S-3)V_{i,l2 \sigma T}  \right) \right]  \vphantom{\frac{1}{2}} \right) + \delta_{1,S} \left( \vphantom{\frac{\sqrt{J}\delta_{J,}}{J}} 2 \delta_{J,l} \delta_{J,l'} \right.  \right. \\ \nonumber 
 - & \ \left. \left.  \frac{2(J-1)\delta_{J-1,l}\delta_{J-1,l'}}{2J+1}-  \frac{2(J+2)\delta_{J+1,l}\delta_{J+1,l'}}{2J+1} + \frac{6\sqrt{J(J+1)}(\delta_{J+1,l}\delta_{J-1,l'}+\delta_{J-1,l}\delta_{J+1,l'})}{2J+1} \right) \left[V_{i,t} + (4T-3)V_{i,t\tau} \right] \right],
 \end{align}
\end{widetext}
where the $tT$, $\tau z$ and $\sigma \tau z$ are not included since in our analysis the contribution of these operators is set to zero. 

Evaluation of equation (\ref{eq:PWOperators}) at the $^1S_0np$,
$^1S_0pp$, $^3P_0$, $^1P_1$, $^3P_1$, $^3S_1$, $\epsilon_1$, $^3D_1$,
$^1D_2$, $^3D_2$, $^3P_2$, $\epsilon_2$, $^3F_2$, $^1F_3$, $^3D_3$
results on a set of fifteen equations. By imposing that the only
charge dependent parameters are on the $^1S_0$ partial wave the
condition $V_{i,T}=-V_{i,\sigma T} = -6 V_{i,l2 T} = 6 V_{i,l2 \sigma
  T} $ must be fulfilled and the number of independent variables is
reduced to fifteen. The solution of this system of equations is
given gy 
\begin{eqnarray}
V_{i,O} = \sum_{JS, l\leq l'} \frac1{M_p \Delta r} c_{O,JSll'} (\lambda_i)^{JS}_{ll'}
\end{eqnarray}
where $O$ runs over the labels of the operator basis. The numerical
values of the $c_{O,JSll'}$ coefficients are displayed in
table~\ref{tab:PW2Operator} where $JSll'$ are expressed in the
spectroscopic notation, $^{2S+1}l_J$ with $l=0,1,2,3,4,5, \dots$
conventionally named $S,P,D,F,G,H, \dots$ for the diagonal matrix
elements. The non-diagonal terms corresponding total angular momentum
$J$ are labeled as $\epsilon_J$.

\begin{table*}
% \footnotesize
 \caption{\label{tab:PW2Operator} Transformation coefficientes $c_{O,JSll'}$
between the fifteen low angular momentum partial waves used to parametrize the $\delta$-shell potential and the fifteen independent operators of the complete potential.}
 \begin{ruledtabular}
 \begin{tabular*}{\textwidth}{@{\extracolsep{\fill}} c *{15}{D{.}{}{-1}}}
 Operator &  \multicolumn{1}{c}{$^1S_0np$} & \multicolumn{1}{c}{$^1S_0pp$} & \multicolumn{1}{c}{$^3P_0$}      & \multicolumn{1}{c}{$^1P_1$} & 
             \multicolumn{1}{c}{$^3P_1$}   & \multicolumn{1}{c}{$^3S_1$}   & \multicolumn{1}{c}{$\epsilon_1$} & \multicolumn{1}{c}{$^3D_1$} & 
             \multicolumn{1}{c}{$^1D_2$}   & \multicolumn{1}{c}{$^3D_2$}   & \multicolumn{1}{c}{$^3P_2$}      & \multicolumn{1}{c}{$\epsilon_2$} &
             \multicolumn{1}{c}{$^3F_2$}   & \multicolumn{1}{c}{$^1F_3$}   & \multicolumn{1}{c}{$^3D_3$} \\
$c$            &  .\frac{1}{16}         &  .\frac{1}{8}   &  .\frac{3}{8}   &  .\frac{3}{40}  &  .\ 0           &  .\frac{3}{16}          &
                  .\ 0                  &  .\ 0           &  .\ 0           &  .\ 0           &  .\frac{3}{10}  &  .\frac{\sqrt{6}}{5}    &
                 -.\frac{9}{80}         & -.\frac{1}{80}  &  .\ 0           \\
$\tau$         &  .\frac{1}{48}         &  .\frac{1}{24}  &  .\frac{1}{8}   & -.\frac{3}{40}  &  .\ 0           & -.\frac{3}{16}          &
                  .\ 0                  &  .\ 0           &  .\ 0           &  .\ 0           &  .\frac{1}{10}  &  .\frac{\sqrt{6}}{15}   &
                 -.\frac{3}{80}         &  .\frac{1}{80}  &  .\ 0           \\
$\sigma$       & -.\frac{1}{16}         & -.\frac{1}{8}   &  .\frac{1}{8}   & -.\frac{3}{40}  &  .\ 0           &  .\frac{1}{16}          &
                  .\ 0                  & .\ 0            &  .\ 0           &  .\ 0           &  .\frac{1}{10}  &  .\frac{\sqrt{6}}{15}   &
                 -.\frac{3}{80}         &  .\frac{1}{80}  &  .\ 0           \\
$\sigma\tau$   & -.\frac{1}{48}         & -.\frac{1}{24}  &  .\frac{1}{24}  &  .\frac{3}{40}  &  .\ 0           & -.\frac{1}{16}          &
                   .\ 0                 & .\ 0            &  .\ 0           &  .\ 0           &  .\frac{1}{30}  &  .\frac{\sqrt{6}}{45}   &
                   -.\frac{1}{80}       & -.\frac{1}{80}  &  .\ 0           \\
$t$            &  .\ 0                  &  .\ 0           &  .\ 0           &  .\ 0           &  .\ 0           &  .\ 0                   &
                  .\frac{\sqrt{2}}{16}  &  .\ 0           &  .\ 0           &  .\ 0           &  .\ 0           &  .\frac{5\sqrt{6}}{48}  &
                  .\ 0                  &  .\ 0           &  .\ 0           \\
$t\tau$        &  .\ 0                  &  .\ 0           &  .\ 0           &  .\ 0           &  .\ 0           &  .\ 0                   &
                 -.\frac{\sqrt{2}}{16}  &  .\ 0           &  .\ 0           &  .\ 0           &  .\ 0           &  .\frac{5\sqrt{6}}{144} &
                  .\ 0                  &  .\ 0           &  .\ 0           \\
$ls$           &  .\ 0                  &  .\ 0           &  .\ 0           &  .\ 0           & -.\frac{3}{8}   &  .\ 0                   &
                  .\frac{\sqrt{2}}{56}  & -.\frac{1}{40}  &  .\ 0           & -.\frac{1}{24}  &  .\frac{3}{8}   &  .\frac{\sqrt{6}}{8}    &
                  .\ 0                  &  .\ 0           &  .\frac{1}{15}  \\
$ls\tau$       &  .\ 0                  &  .\ 0           &  .\ 0           &  .\ 0           & -.\frac{1}{8}   &  .\ 0                   &
                 -.\frac{\sqrt{2}}{56}  & .\frac{1}{40}   &  .\ 0           &  .\frac{1}{24}  &  .\frac{1}{8}   &  .\frac{\sqrt{6}}{24}   &
                  .\ 0                  &  .\ 0           & -.\frac{1}{15}  \\
$l2$           & -.\frac{1}{96}         & -.\frac{1}{48}  & -.\frac{9}{32}  & -.\frac{1}{160} &  .\frac{9}{32}  & -.\frac{1}{32}          &
                 -.\frac{\sqrt{2}}{56}  & -.\frac{1}{160} &  .\frac{1}{32}  &  .\frac{1}{32}  & -.\frac{9}{160} & -.\frac{9\sqrt{6}}{40}  &
                  .\frac{9}{160}        &  .\frac{1}{160} &  .\frac{1}{160} \\
$l2\sigma$     &  .\frac{1}{96}         &  .\frac{1}{48}  & -.\frac{3}{32}  &  .\frac{1}{160} &  .\frac{3}{32}  & -.\frac{1}{96}          &
                 -.\frac{\sqrt{2}}{168} & -.\frac{1}{480} & -.\frac{1}{32}  &  .\frac{1}{96}  & -.\frac{3}{160} & -.\frac{3\sqrt{6}}{40}  &
                  .\frac{3}{160}        & -.\frac{1}{160} &  .\frac{1}{480} \\
$l2\tau$       & -.\frac{1}{288}        & -.\frac{1}{144} & -.\frac{3}{32}  &  .\frac{1}{160} &  .\frac{3}{32}  &  .\frac{1}{32}          &
                  .\frac{\sqrt{2}}{56}  &  .\frac{1}{160} &  .\frac{1}{96}  & -.\frac{1}{32}  & -.\frac{3}{160} & -.\frac{3\sqrt{6}}{40}  &
                  .\frac{3}{160}        & -.\frac{1}{160} & -.\frac{1}{160} \\
$l2\sigma\tau$ &  .\frac{1}{288}        &  .\frac{1}{144} & -.\frac{1}{32}  & -.\frac{1}{160} &  .\frac{1}{32}  &  .\frac{1}{96}          &
                  .\frac{\sqrt{2}}{168} &  .\frac{1}{480} & -.\frac{1}{96}  & -.\frac{1}{96}  & -.\frac{1}{160} & -.\frac{\sqrt{6}}{40}   &
                  .\frac{1}{160}        &  .\frac{1}{160} & -.\frac{1}{480} \\
$ls2$          &  .\ 0                  &  .\ 0           &  .\frac{1}{4}   &  .\ 0           & -.\frac{3}{8}   &  .\ 0                   &
                  .\frac{\sqrt{2}}{28}  &  .\frac{1}{40}  &  .\ 0           & -.\frac{1}{24}  &  .\frac{1}{8}   &  .\frac{\sqrt{6}}{4}    &
                  .\ 0                  &  .\ 0           &  .\frac{1}{60}  \\
$ls2\tau$      &  .\ 0                  &  .\ 0           &  .\frac{1}{12}  &  .\ 0           & -.\frac{1}{8}   &  .\ 0                   &
                 -.\frac{\sqrt{2}}{28}  & -.\frac{1}{40}  &  .\ 0           &  .\frac{1}{24}  &  .\frac{1}{24}  &  .\frac{\sqrt{6}}{12}   &
                  .\ 0                  &  .\ 0           & -.\frac{1}{60}  \\
$T$            & -.\frac{1}{24}         &  .\frac{1}{24}  &  .\ 0           &  .\ 0           &  .\ 0           &  .\ 0                   &
                  .\ 0                  &  .\ 0           &  .\ 0           &  .\ 0           &  .\ 0           &  .\ 0                   &
                  .\ 0                  &  .\ 0           &  .\ 0           \\
 \end{tabular*}
 \end{ruledtabular}
\end{table*}

\section{Calculation of phase shifts from $\delta$-shell potential}
\label{sec:sch}
\subsection{Uncoupled partial waves}
\label{subsec:CentralP}
 The two body scattering amplitued $M$ can be constructed directly using the phaseshift for every partial wave. To calculate such phaseshifts  it is necessary to integrate the corresponding Schr\"odinger equation. In the case of uncoupled partial waves ($l=l'=J$) the potential is central and the reduced  equation reads
 \begin{equation}
  \label{eq:ReducedSchr}
  - \frac{d^2}{d r^2} u_l(k,r)+ \left[ U_{l,l}^{l,S}(r) + \frac{l(l+1)}{r^2} \right] u_l(k,r) = k^2 u_l(k,r),
 \end{equation}
where $U_{l,l}^{J,S}(r) = 2 \mu_{\alpha,\beta}V_{l,l}^{J,S}(r)$ is the reduced potential. To solve this differential equation we use the fact that inside the region $r_{i} < r < r_{i+1}$ the $\delta$-shell potential is zero and $u_l(k,r) = u_{l,i} (k,r)$, with $u_{l,i}$ the solution of the free particle equation
 \begin{equation}
 \label{eq:SchrFreeParticle}
  -\frac{d^2}{d r^2} u_{l,i}(k,r)+  \frac{l(l+1)}{r^2}  u_{l,i}(k,r) = k^2 u_{l,i}(k,r).
  \end{equation}
Therefore in the neighborhood of $r_i$ we can take a solution of the type
 \begin{equation}
  \label{eq:SolNeigborRi}
  u_{l}(k,r)=[1- \theta(r-r_i)]u_{l,i-1}(k,r) + \theta(r-r_i) u_{l,i}(k,r).
 \end{equation}
 
Substituting this solution in Eq.~(\ref{eq:ReducedSchr}) and considering the linear independence of $\delta(r-r_i)$ and $\delta'(r-r_i)$, two conditions are found to hold at every concentration radius
 \begin{eqnarray}
  u_{l,i}(k,r_i) = u_{l,i-1}(k,r_i) &\equiv& u(k,r_i), \nonumber \\
  \label{eq:BoundaryCondUncoupled}
  \left[ \frac{d}{d r} u_{l,i}(k,r) - \frac{d}{d r} u_{l,i-1}(k,r) \right]_{r=r_i}  &=& \lambda_i u(k,r_i),
 \end{eqnarray}
where the $l,S,J$ labels of the $\lambda_i$ coefficients have been omitted. The first equation implies the continuity of the wave function, while the second expresses the relation between the strength coefficients $\lambda_i$ and the change in the log derivative of wave function at the interaction points $r_i$.

Since $u_{l,i}(k,r)$ is the solution of the free particle equation, it can be written as a linear combination of the reduced spherical Bessel functions
\begin{eqnarray}
\label{eq:ReducedBessel}
\hat{j}_l(x) = x j_l(x), \ \ \hat{y}_l(x) = x y_l(x).
\end{eqnarray}
For convenience we choose a linear combination with the form
\begin{equation}
\label{eq:UiCombBessel}
u_{l,i}(k,r)=A_i\left[ \hat{j}_l(k r) - \tan{\delta_i(k)} \hat{y}_l(k r) \right],
\end{equation} 
where $A_i$ and $\delta_i$ are constants to be determined.
This representation allows to take a wave function with the asymptotic form
\begin{equation}
\label{eq:AsymptoticWF}
u_{l,N}(k,r)= \hat{j}_l(kr) - \tan{\delta_N} \hat{y}_l(kr)
\end{equation}
for distances greater than the last concentration radius $r_N$. In this case $\delta_N$ is the phase shift resulting from the scattering process. The condition of regularity at the origin is met with
 \begin{equation}
  \label{eq:u0RegularOrigen}
  u_{l,0}(k,r) = A_0 \hat{j}_l (k r),
 \end{equation}
which imposes the condition $\tan{\delta_0}=0$.

Now, only an equation for expressing $\tan{\delta_i}(k)$ in terms of $\tan{\delta_{i-1}}(k)$ is needed to calculate $\delta_N(k)$. For simplicity we define
\begin{equation}
\label{eq:WFphi}
\varphi_{l,i}(k,r)=\hat{j}_l(kr)-\tan{\delta_i(k)}\hat{y}_l(k r),
\end{equation}
which allows us to write
\begin{equation}
\label{eq:uPhi}
u_{l}(k,r_i) = A_i \varphi_{l,i}(k,r_i) = A_{i-1} \varphi_{l,i-1}(k,r_i).
\end{equation}
Multiplying the second equation in (\ref{eq:BoundaryCondUncoupled}) by $u_l(k,r)$ and using (\ref{eq:uPhi}) conveniently the term $A_{i-1} A_i$ appears on both sides of the equation and we get
\begin{eqnarray}
  \varphi_{l,i-1}(k,r_i)&&\left. \frac{d}{d r} \varphi_{l,i}(k,r)   \right|_{r=r_i}  \nonumber \\
 &&-\varphi_{l,i}(k,r_i)   \left. \frac{d}{d r} \varphi_{l,i-1}(k,r) \right|_{r=r_i}  \nonumber \\
 & &\label{eq:CabioVarphi}
 =\lambda_i \varphi_{l,i-1}(k,r_i)\varphi_{l,i}(k,r_i). 
\end{eqnarray}
Inserting the definition (\ref{eq:WFphi}) and considering the Wronskian relation
\begin{equation}
\label{eq:Wronskian}
\hat{y}_l(k r) \frac{d}{dr} \hat{j}_l(kr) - \hat{j}_l(k r) \frac{d}{dr} \hat{y}_l(kr ) = k
\end{equation}
we get 
\begin{equation}
\label{eq:DifTanDeltai}
\tan{\delta_i(k)}-\tan{\delta_{i-1}(k)}=- \frac{\lambda_i}{k} \varphi_{l,i-1}(k,r_i)\varphi_{l,i}(k,r_i).
\end{equation}
This expression can be considered a discrete version of the variable phase equation, since in the limit of a continuous interaction  the form of the latter is recovered. Finally, solving for $\tan{\delta_i(k)}$ we get
\begin{equation}
\label{eq:tDeltai}
\tan{\delta_i(k)}= \frac{\tan{\delta_{i-1}(k)}-\frac{\lambda_i}{k}\hat{j}_l(kr_i)\varphi_{l,i-1}(k,r_i)}{1-\frac{\lambda_i}{k}\hat{y}_l(kr_i)\varphi_{l,i-1}(k,r_i)}
\end{equation}
and using  $\tan{\delta_0}=0$ as a boundary condition we are able to calculate $\delta_N$.

For completeness we show how to get the entire solution to (\ref{eq:ReducedSchr}). Eq.~(\ref{eq:AsymptoticWF}) implies the condition $A_N=1$, this allows to calculate all the $A_i$ constants using the equation (\ref{eq:uPhi}) as
\begin{equation}
\label{eq:Ai1Ai}
A_{i-1} = A_i \varphi_{l,i}(k,r_i)\varphi_{l,i-1}^{-1}(k,r_i).
\end{equation}

\subsection{Coupled partial waves}
\label{sec:TensorP}
The tensor part of the NN interaction couples the triplet partial waves with $l=l'=J\pm1$. In this case a set of two differential equations must be solved simultaniously to extract the corresponding phaseshifts. The coupled reduced Schr\"odinger equation reads
\begin{widetext}
\begin{eqnarray}
-\frac{d^2}{dr^2}v(k,r)+\left[ U_{J-1,J-1}^{J,1}(r) + \frac{(J-1)J}{r^2} \right] v(k,r) +U_{J-1,J+1}^{J,1}(r) w(k,r) &=& k^2 v(k,r) \nonumber \\
\label{eq:SchroTripletCoup}
-\frac{d^2}{dr^2}w(k,r)+\left[ U_{J+1,J+1}^{J,1}(r) + \frac{(J+1)(J+2)}{r^2} \right] w(k,r) +U_{J+1,J-1}^{J,1}(r) v(k,r) &=& k^2 w(k,r) 
\end{eqnarray}
\end{widetext}
This equation has two linearly independent solutions that we can label $\alpha$ and $\beta$, and its' asymptotic behavior can be written as 
\begin{eqnarray}
v^{(\alpha)}(k,r)&\rightarrow&\cot{\delta_{J-1}^1}\hat{j}_{J-1}(kr)  -  \hat{y}_{J-1}(kr) \nonumber  \\
w^{(\alpha)}(k,r)&\rightarrow&\tan{\epsilon_J} [ \cot{\delta_{J-1}^1}\hat{j}_{J+1}(kr)  -  \hat{y}_{J+1}(kr)] \nonumber  \\
&& \nonumber \\
v^{(\beta)}(k,r)&\rightarrow&-\tan{\epsilon_J} [ \cot{\delta_{J+1}^1}\hat{j}_{J-1}(kr)  -  \hat{y}_{J-1}(kr)] \nonumber  \\
\label{eq:AsymptoticSchrCoupledSol}
w^{(\beta)}(k,r)&\rightarrow&\cot{\delta_{J+1}^1}\hat{j}_{J+1}(kr)  -  \hat{y}_{J+1}(kr)
\end{eqnarray}
where $\delta_{J-1}$, $\delta_{J+1}$ and $\epsilon_J$ are the phase shifts in the BB or Eigen phase parametrization \cite{Blatt:1952zza}.
For a given value of $J$ we use the following notation for the $\delta$-shell reduced potential matrix elements
\begin{eqnarray}
U_{J-1,J-1}^{J,1}&=&\sum_{i=1}^N{\lambda_{i}^{J-1}\delta(r-r_i)},  \nonumber \\
U_{J+1,J+1}^{J,1}&=&\sum_{i=1}^N{\lambda_{i}^{J+1}\delta(r-r_i)}, \nonumber \\
\label{eq:CoupledPotentialDeltaShell}
U_{J-1,J+1}^{J,1} = U_{J+1,J-1}^{J,1} &=& \sum_{i=1}^N{\tilde{\lambda}_{i}\delta(r-r_i)}.
\end{eqnarray}

Similarly as we did for the central potential case, we consider the solution inside the interval $r_i < r < r_{i+1}$ to be the solution of the free particle equations
\begin{eqnarray}
-\frac{d^2}{dr^2}v_i(k,r)+ \frac{(J-1)J}{r^2} v_i(k,r) &=& k^2 v_i(k,r), \nonumber \\
-\frac{d^2}{dr^2}w_i(k,r)+\frac{(J+1)(J+2)}{r^2} w_i(k,r) &=& k^2 w_i(k,r). \nonumber \\
\label{eq:SchrFreeTriplet}
\end{eqnarray}
Again, we construct the solution in the neighborhood of $r_i$ as
\begin{eqnarray}
v(k,r)&=&[1-\theta(r-r_i)]v_{i-1}(k,r)+\theta(r-r_i)v_{i}(k,r),  \nonumber \\
w(k,r)&=&[1-\theta(r-r_i)]w_{i-1}(k,r)+\theta(r-r_i)w_{i}(k,r). \nonumber \\
\label{eq:CoupledFreeParticle}
\end{eqnarray}
Substituting these solutions in Eq.~(\ref{eq:SchroTripletCoup}) with the reduced $\delta$-shell potential it is possible to get boundary conditions similar to the ones in (\ref{eq:BoundaryCondUncoupled})
\begin{eqnarray}
v_{i}(k,r_i)=v_{i-1}(k,r_i)&\equiv&v(k,r_i), \nonumber \\
w_{i}(k,r_i)=w_{i-1}(k,r_i)&\equiv&w(k,r_i), \nonumber \\
\left[\frac{d}{dr}v_{i}(k,r) - \frac{d}{dr}v_{i-1}(k,r) \right]_{r=r_i}&=&\lambda_{i}^{J-1} v(k,r_i) \nonumber \\ 
&+& \tilde{\lambda}_{i} w(k,r_i), \nonumber \\
\left[\frac{d}{dr}w_{i}(k,r) - \frac{d}{dr}w_{i-1}(k,r) \right]_{r=r_i}&=&\lambda_{i}^{J+1} w(k,r_i) \nonumber \\ 
\label{eq:BoundaryCondCoupled}
&+& \tilde{\lambda}_{i} v(k,r_i).
\end{eqnarray}

To integrate the coupled Schr\"odinger equation we consider two linearly independent solutions $(v_1,w_1)$ and $(v_2,w_2)$ that in the region $0 \leq r < r_1 $ are
\begin{eqnarray}
v_{1,0}(k,r)=\hat{j}&_{J-1}(kr), \ \ \  &w_{1,0}(k,r)=0, \nonumber \\
\label{eq:CoupledSolr0}
v_{2,0}(k,r)=0,& \ \ \  &w_{2,0}(k,r)=\hat{j}_{J+1}(kr).
\end{eqnarray}
Some explanation should be given to the indices, the first one is used to differentiate between the two linearly independent solutions while the second indicates free particle solution inside the interval $r_{i} < r < r_{i+1}$.
For consistency in notation we consider $r_0 = 0$ but it should be noted that no concentration radius at $r=0$ is present in the delta shell potentials of Eq.~(\ref{eq:CoupledPotentialDeltaShell}).

Also should be pointed out that the differential equations in (\ref{eq:SchrFreeTriplet}) are uncoupled and therefore the solutions can be expressed again as a linear combination of the reduced spherical bessel functions.
With this in mind we write
\begin{eqnarray}
\label{eq:FreeCoupledI}
v_{1,i}(k,r)&=&A_{1,i}\hat{j}_{J-1}(kr) + B_{1,i}\hat{y}_{J-1}(kr), \nonumber \\
w_{1,i}(k,r)&=&C_{1,i}\hat{j}_{J+1}(kr) + D_{1,i}\hat{y}_{J+1}(kr), \nonumber \\
\nonumber \\
v_{2,i}(k,r)&=&A_{2,i}\hat{j}_{J-1}(kr) + B_{2,i}\hat{y}_{J-1}(kr), \nonumber \\
w_{2,i}(k,r)&=&C_{2,i}\hat{j}_{J+1}(kr) + D_{2,i}\hat{y}_{J+1}(kr).
\end{eqnarray}
The $\alpha$ and $\beta$ asymptotic wave functions can be formed as  linear combinations of both solutions for distances greater than the last concentration radius $r_N$, i.e.
\begin{eqnarray}
\label{eq:AsymptoticLinearCombCoupled}
v^{\alpha}(k,r)&=&\alpha_1 v_{1,N}(k,r) + \alpha_2 v_{2,N}(k,r),  \nonumber \\
w^{\alpha}(k,r)&=&\alpha_1 w_{1,N}(k,r) + \alpha_2 w_{2,N}(k,r),  \nonumber \\
\nonumber \\
v^{\beta}(k,r)&=&\beta_1 v_{1,N}(k,r) + \beta_2 v_{2,N}(k,r),  \nonumber \\
w^{\beta}(k,r)&=&\beta_1 w_{1,N}(k,r) + \beta_2 w_{2,N}(k,r).  
\end{eqnarray}
Matching this linear combinations in the form of Eq.~(\ref{eq:FreeCoupledI}) with the asymptotic solutions of Eq.~(\ref{eq:AsymptoticSchrCoupledSol}) it is possible to get the expressions
\begin{eqnarray}
\label{eq:alphaPhaseShifts}
\frac{A_{1,N}+\alpha A_{2_N}}{B_{1,N}+\alpha B_{2,N}}&=&\frac{C_{1,N}+\alpha C_{2,N}}{D_{1,N}+\alpha D_{2,N}} = \cot{\delta_{J-1(J+1)}^1}, \nonumber  \\
\frac{D_{1,N}+\alpha D_{2,N}}{B_{1,N}+\alpha B_{2,N}}&=& \tan{\epsilon_J}
\end{eqnarray}
where we have defined $\alpha \equiv \alpha_2/\alpha_1$.
The first equation has two solutions on $\alpha$, one corresponds to the $\alpha$-state eigen-phase shift and the other to the $\beta$ one. The second equation can be used unambiguously to obtain the mixing angle $\epsilon_J$. 

Matching the equations of Eq.~(\ref{eq:CoupledSolr0}) with the ones in (\ref{eq:FreeCoupledI}) we can get
\begin{eqnarray}
\label{eq:ABCD0}
A_{1,0} = 1,& \ \ \ & B_{1,0} = 0, \nonumber \\
C_{1,0} = 0,&  & D_{1,0} = 0, \nonumber \\
A_{2,0} = 0,&  & B_{2,0} = 0, \nonumber \\
C_{2,0} = 1,&  & D_{2,0} = 0. 
\end{eqnarray}
Like in the previous subsection, we need an expression for the $v_{1(2),i},w_{1(2),i}$ wavefunctions in terms of the $v_{1(2),i-1},w_{1(2),i-1}$ ones.
Combining the boundary conditions for (\ref{eq:BoundaryCondCoupled}) with the linear combinations of (\ref{eq:FreeCoupledI}) and using properly the Wronskian relation in Eq.~(\ref{eq:Wronskian}) we were able to get following expressions for the constants $\{A,B,C,D\}_{1(2),i}$ in terms of the constants $\{A,B,C,D\}_{1(2),i-1}$
\begin{widetext}
\begin{eqnarray}
\label{eq:ABCDiABCDim1}
B_{i}&=&B_{i-1}+\frac{1}{k} \hat{j}_{J-1}(kr_i) \{\lambda_i^{J-1} [A_{i-1} \hat{j}_{J-1}(k r_i) + B_{i-1}\hat{y}_{J-1}(k r_i)] + \tilde{\lambda}_i [C_{i-1}\hat{j}_{J+1}(kr_i) + D_{i-1}\hat{y}_{J+1}(k r_i)] \}, \nonumber   \\
A_{i}&=& [\hat{j}_{J-1}(kr_i)]^{-1}[A_{i-1}\hat{j}_{J-1}(kr_i) + (B_{i-1}-B_i) \hat{y}_{J-1}(kr_i)], \nonumber \\
D_{i}&=&D_{i-1}+\frac{1}{k} \hat{j}_{J+1}(kr_i) \{\lambda_i^{J+1} [C_{i-1} \hat{j}_{J+1}(k r_i) + D_{i-1}\hat{y}_{J+1}(k r_i)] + \tilde{\lambda}_i [A_{i-1}\hat{j}_{J-1}(kr_i) + B_{i-1}\hat{y}_{J-1}(k r_i)] \}, \nonumber   \\
C_{i}&=& [\hat{j}_{J+1}(kr_i)]^{-1}[C_{i-1}\hat{j}_{J+1}(kr_i) + (D_{i-1}-D_i) \hat{y}_{J+1}(kr_i)] 
\end{eqnarray}
\end{widetext}
where the $1$ and $2$ indices have been suppressed for simplicity.

Finally, the relations in Eq.~(\ref{eq:ABCDiABCDim1}) can be used with the boundary conditions of (\ref{eq:ABCD0}) to calculate the constants $\{ A,B,C,D \}_{1(2),N}$ and ultimately use them together with (\ref{eq:alphaPhaseShifts}) to get the Eigen-phase shifts for the coupled channels with orbital angular momentum $l=J\pm1$.

\section{Numerical details}

As already noted in our previous works, the $\delta$-shell
representation is just a method to solve Schr\"odinger's equation.
Thus it can also be used to integrate the long range part of the
interaction such as OPE. The only difference is that for pp the free
particle wave functions are replaced by both regular and irregular
Coulomb wave functions. Some caution is needed handling the boundary
$r_i=r_c$ since Coulomb wave functions from the $r> r_c$ region must
match the free particle wave functions for $r< r_c$.  In this case the
$\delta$-shell coefficients are {\it fixed} to the long range
potential value, i.e. $V_i \equiv V(r_i)$, and are no longer fitting
parameters. Therefore, the accuracy of the integration depends on the
number of sampling points for $r > r_c$.  According to our discussion
in Refs.~\cite{NavarroPerez:2011fm,NavarroPerez2013138,Perez:2013mwa}
Nyquist optimal sampling theorem suggests keeping $\Delta r=0.6 {\rm
  fm}$ throughout.  For OPE it is sufficient to take $N \sim 20$.  As
a final remark, note that the question of accuracy in the unknown
region never arises, i.e. there is no point in improving {\it beyond}
the $\Delta r \sim 1/p_{\rm max}$ resolution for a maximum CM momentum
$p_{\rm max}$ which in our case is given by pion production threshold. 

%\bibliography{deltashell}

%Merlin.mbs v4.21 2009-07-09.
%

\end{document}

%% file: keptpp.tex
  0.3 -   0.4 & BR64\cite{Brolley:1964zz}     & DSG  &      91.2     &   5 &   5 & \multicolumn{2}{c}{Float} & 0.735        &    8.84 &  1.77 \\
  0.4         & TH78\cite{Thomann1978457}     & DSG  &  58.4 -  95.0 &   4 &   4 & \multicolumn{2}{c}{Float} & 1.003        &    0.76 &  0.19 \\
  0.4         & TH78\cite{Thomann1978457}     & DSG  &  26.0 - 110.0 &  17 &  18 &  0.2\% & 0.024            & 1.000        &   32.72 &  1.82 \\
  0.4         & TH78\cite{Thomann1978457}     & DSG  &  58.4 -  95.0 &   4 &   4 & \multicolumn{2}{c}{Float} & 0.999        &   10.94 &  2.74 \\
  0.4         & TH78\cite{Thomann1978457}     & DSG  &  58.4 -  95.0 &   4 &   4 & \multicolumn{2}{c}{Float} & 1.000        &    5.62 &  1.40 \\
  0.4         & TH78\cite{Thomann1978457}     & DSG  &  58.4 -  98.0 &   6 &   6 & \multicolumn{2}{c}{Float} & 0.999        &   16.39 &  2.73 \\
  0.4         & TH78\cite{Thomann1978457}     & DSG  &  58.4 -  95.0 &   4 &   4 & \multicolumn{2}{c}{Float} & 0.999        &   13.01 &  3.25 \\
  0.4         & TH78\cite{Thomann1978457}     & DSG  &  40.0 -  46.0 &   3 &   4 &  0.2\% & 0.094            & 0.999        &    0.97 &  0.24 \\

%% file: keptnp.tex
  0.0         & LO74\cite{PhysRevC.9.1329}      & SGT  &       -       &   1 &   1 & \multicolumn{3}{c}{No systematic error}  &    2.95 &  2.95 \\
  0.0         & DI75\cite{PhysRevC.11.103}      & SGT  &       -       &   1 &   1 & \multicolumn{3}{c}{No systematic error}  &    4.23 &  4.23 \\
  0.0         & HO71\cite{PhysRevC.3.1886}      & SGT  &       -       &   1 &   1 & \multicolumn{3}{c}{No systematic error}  &    0.37 &  0.37 \\
  0.0         & KO90\cite{KoesterZPA272.2.189}  & SGT  &       -       &   1 &   1 & \multicolumn{3}{c}{No systematic error}  &    8.04 &  8.04 \\
  0.0         & FU76\cite{Fujita19761}          & SGT  &       -       &   1 &   1 & \multicolumn{3}{c}{No systematic error}  &    4.27 &  4.27 \\
  0.1 -   0.6 & AL55\cite{AllenPPSA68.11.1077}  & SGT  &       -       &   5 &   5 & \multicolumn{3}{c}{No systematic error}  &    2.26 &  0.45 \\
  0.2 -   0.8 & DA13\cite{PhysRevC.87.014005}   & SGT  &       -       &  65 &  65 & \multicolumn{3}{c}{No systematic error}  &   90.36 &  1.39 \\
  0.5 -   3.2 & EN63\cite{PhysRev.129.324}      & SGT  &       -       &   2 &   2 & \multicolumn{3}{c}{No systematic error}  &    3.97 &  1.99 \\